\documentclass[10pt, conference, letterpaper, nofonttune]{IEEEtran}
 
\usepackage{url}  
\usepackage{eso-pic}
\usepackage{graphicx}
\usepackage{tikz}
\usepackage{pgf}
\usepackage{lscape}
\usepackage{amssymb}
\usepackage{fancyvrb}
\usepackage{amsfonts}
\usepackage{verbatim}
\usepackage{xfrac}
\usepackage{algorithmic}
\usepackage[draft]{hyperref}
\usepackage{amsmath}
\usepackage{multirow}
\usepackage{algorithm}
\usepackage{proof}
\usepackage{epstopdf}
\usepackage{subcaption}
\usepackage[nolist,nohyperlinks]{acronym}
\usepackage[acronym]{glossaries}
\usepackage[belowskip=-5pt, font=small]{caption}
\usepackage[numbers,sort&compress]{natbib}
\usepackage{xspace}
\usepackage{balance}

\acrodef{arq}[ARQ]{Automatic Repeat Request}
\acrodef{ask}[ASK]{Amplitude Shift Keying}
\acrodef{bs}[BS]{Base Station}
\acrodef{ccaas}[CCaaS]{Cellular Connectivity as a Service}
\acrodef{cdf}[CDF]{Cumulative Distribution Function}
\acrodef{cots}[COTS]{Commercial Off-the-Shelf}
\acrodef{crc}[CRC]{Cyclic Redundancy Check}
\acrodef{dlsch}[DL-SCH]{Downlink Shared Channel}
\acrodef{epc}[EPC]{Evolved Packet Core}
\acrodef{hmac}[HMAC]{Keyed-Hash Message Authentication Code}
\acrodef{hss}[HSS]{Home Subscriber Server}
\acrodef{icmp}[ICMP]{Internet Control Message Protocol}
\acrodef{imsi}[IMSI]{International Mobile Subscriber Identity}
\acrodef{iot}[IoT]{Internet of Things}
\acrodef{ip}[IP]{Infrastructure Provider}
\acrodef{ks}[K-S]{Kolmogorov–Smirnov}
\acrodef{mac}[MAC]{Medium Access Control}
\acrodef{mme}[MME]{Mobility Management Entity}
\acrodef{msin}[MSIN]{Mobile Subscription Identification Number}
\acrodef{mvno}[MVNO]{Mobile Virtual Network Operator}
\acrodef{nbiot}[NB-IoT]{Narrowband \ac{iot}}
\acrodef{pawr}[PAWR]{Platforms for Advanced Wireless Research}
\acrodef{pccaas}[PCCaaS]{Private Cellular Connectivity as a Service}
\acrodef{pdf}[PDF]{Probability Density Function}
\acrodef{pdsch}[PDSCH]{Physical Downlink Shared Channel}
\acrodef{pgw}[PGW]{Packet Data Network Gateway}
\acrodef{powder}[POWDER]{Platform for Open Wireless Data-driven Experimental Research}
\acrodef{prb}[PRB]{Physical Resource Block}
\acrodef{pusch}[PUSCH]{Physical Uplink Shared Channel}
\acrodef{qoe}[QoE]{Quality of Experience}
\acrodef{rach}[RACH]{Random Access Channel}
\acrodef{ran}[RAN]{Radio Access Network}
\acrodef{rf}[RF]{Radio Frequency}
\acrodef{sdr}[SDR]{Software-defined Radio}
\acrodef{sgw}[SGW]{Serving Gateway}
\acrodef{sinr}[SINR]{Signal-to-Interference-plus-Noise Ratio}
\acrodef{ue}[UE]{User Equipment}
\acrodef{ulsch}[UL-SCH]{Uplink Shared Channel}

\newcommand{\stealte}{\textit{SteaLTE}\xspace}

\newcommand{\fig}[1]{Fig.~\ref{#1}}

\newcommand{\enb}{eNB\xspace}
\newcommand{\enbs}{eNBs\xspace}
\newcommand{\bs}{\ac{bs}\xspace}
\newcommand{\gnb}{gNB\xspace}
\newcommand{\ue}{\ac{ue}\xspace}
\newcommand{\ues}{\acp{ue}\xspace}
\newcommand{\eg}{e.g.,\xspace}
\newcommand{\ie}{i.e.,\xspace}
\newcommand{\ack}{\texttt{ACK}\xspace}
\newcommand{\nack}{\texttt{NACK}\xspace}
\newcommand{\bask}{$2$-ASK\xspace}
\newcommand{\qask}{$4$-ASK\xspace}
\newcommand{\ccaas}{\ac{ccaas}\xspace}
\newcommand{\pccaas}{\ac{pccaas}\xspace}

\title{\textit{SteaLTE}: Private 5G Cellular Connectivity as a Service with Full-stack Wireless Steganography}

\author{\IEEEauthorblockN{Leonardo Bonati,\IEEEauthorrefmark{1} 
Salvatore D'Oro,\IEEEauthorrefmark{1} 
Francesco Restuccia,\IEEEauthorrefmark{1}\IEEEauthorrefmark{2} 
Stefano Basagni,\IEEEauthorrefmark{1} 
Tommaso Melodia\IEEEauthorrefmark{1}}
\IEEEauthorblockA{
\IEEEauthorrefmark{1} Institute for the Wireless Internet of Things, Northeastern University, Boston, MA, U.S.A. \\
\IEEEauthorrefmark{2} Roux Institute, Northeastern University, Portland, ME, U.S.A. \\
\vspace{-1.5cm}}
\thanks{E-mail: \{l.bonati, s.doro, frestuc, s.basagni, t.melodia\}@northeastern.edu}
\thanks{This work was supported by the U.S.\ Office of Naval Research under Grants N00014-19-1-2409 and N00014-20-1-2132.}
}

\IEEEoverridecommandlockouts

\begin{document}

\maketitle

\begin{abstract} 
Fifth-generation (5G) systems will extensively employ radio access network (RAN) softwarization. 
This key innovation enables the instantiation of ``virtual cellular networks'' running on different \emph{slices} of the shared physical infrastructure. 
In this paper, we propose the concept of \textit{Private Cellular Connectivity as a Service} (PCCaaS), where infrastructure providers deploy \textit{covert network slices} known only to a subset of users. 
%
%
We then present \stealte as the first realization of a PCCaaS-enabling system for cellular networks.
At its core, \stealte utilizes \textit{wireless steganography} to disguise data as noise to adversarial receivers. 
Differently from previous work, however, it takes a \textit{full-stack approach} to steganography, 
contributing
%
an LTE-compliant steganographic protocol stack for PCCaaS-based communications, and 
%
packet schedulers and operations to embed \textit{covert data streams} on top of traditional cellular traffic (\textit{primary traffic}). 
\stealte balances undetectability and performance by mimicking channel impairments so that covert data waveforms are almost indistinguishable from noise. 
We evaluate the performance of \stealte on an indoor LTE-compliant testbed under different traffic profiles, distance and mobility patterns. 
We further test it on the outdoor PAWR POWDER platform over long-range cellular links.
Results show that in most experiments \stealte imposes little loss of primary traffic throughput in presence of covert data transmissions ($< 6\%$), making it suitable for undetectable PCCaaS networking.
\end{abstract}

\begin{IEEEkeywords} 
Steganography, 5G, Private Cellular Connectivity as a Service, Undetectability.
\end{IEEEkeywords}

\begin{picture}(0,0)(10,-410)
\put(0,0){
\put(0,10){\footnotesize This paper has been accepted for publication at IEEE International Conference on Computer Communications (INFOCOM) 2021.}
\put(0,0){\tiny \copyright 2021 IEEE. Personal use of this material is permitted. Permission from IEEE must be obtained for all other uses, in any current or future media including reprinting/republishing}
\put(0,-5){\tiny this material for advertising or promotional purposes, creating new collective works, for resale or redistribution to servers or lists, or reuse of any copyrighted component of this work in other works.}
\put(0,-20){\scriptsize }}
\end{picture}

\section{Introduction}

The  \emph{softwarization} of the \ac{ran} is being heralded as the core of fifth generation (5G) cellular networks~\cite{rost2017network, oran2018oran, linux2018onap, moradi2014softmow, bonati2020open, bonati2020intelligence, bonati2020cellos}.
Enabling virtualization technologies, softwarization will allow \acp{ip} to create \textit{virtual networks} on top of their physical infrastructure, each assigned to a different infrastructure \emph{slice}~\cite{doroInfocom2019Slicing, doro2020slicing, doro2020sledge}.
This fundamental innovation will concretely realize the long-standing vision of \textit{\ccaas}, where the \ac{ip} assigns physical resources (\eg spectrum,  power, base stations, etc.) to each \ac{mvno} according to their requirements~\cite{afolabi2018network, marquez2019resource}. 
\ccaas is envisioned to provide unparalleled levels of \ac{qoe} to mobile users, as well as usher in new business opportunities between \acp{ip} and \acp{mvno}~\cite{bega2017optimising,foukas2017network}. %

In this paper we leverage \ac{ran} softwarization and network slicing 
to concretely realize \textit{\pccaas}, pushing the \ccaas innovation to the realm of \emph{private} networking.
Through \pccaas the \acp{ip} can instantiate and deploy \emph{private network slices} sharing the virtualized infrastructure with other (public) slices. 
In this paper we use the word \textit{private} to identify slices whose existence is known only to selected users that can exchange sensitive data embedding it \emph{covertly and undetectably} into \emph{primary traffic}, used as decoy. 

\begin{figure}[h]
    \begin{center}
        \includegraphics[width=0.9\columnwidth]{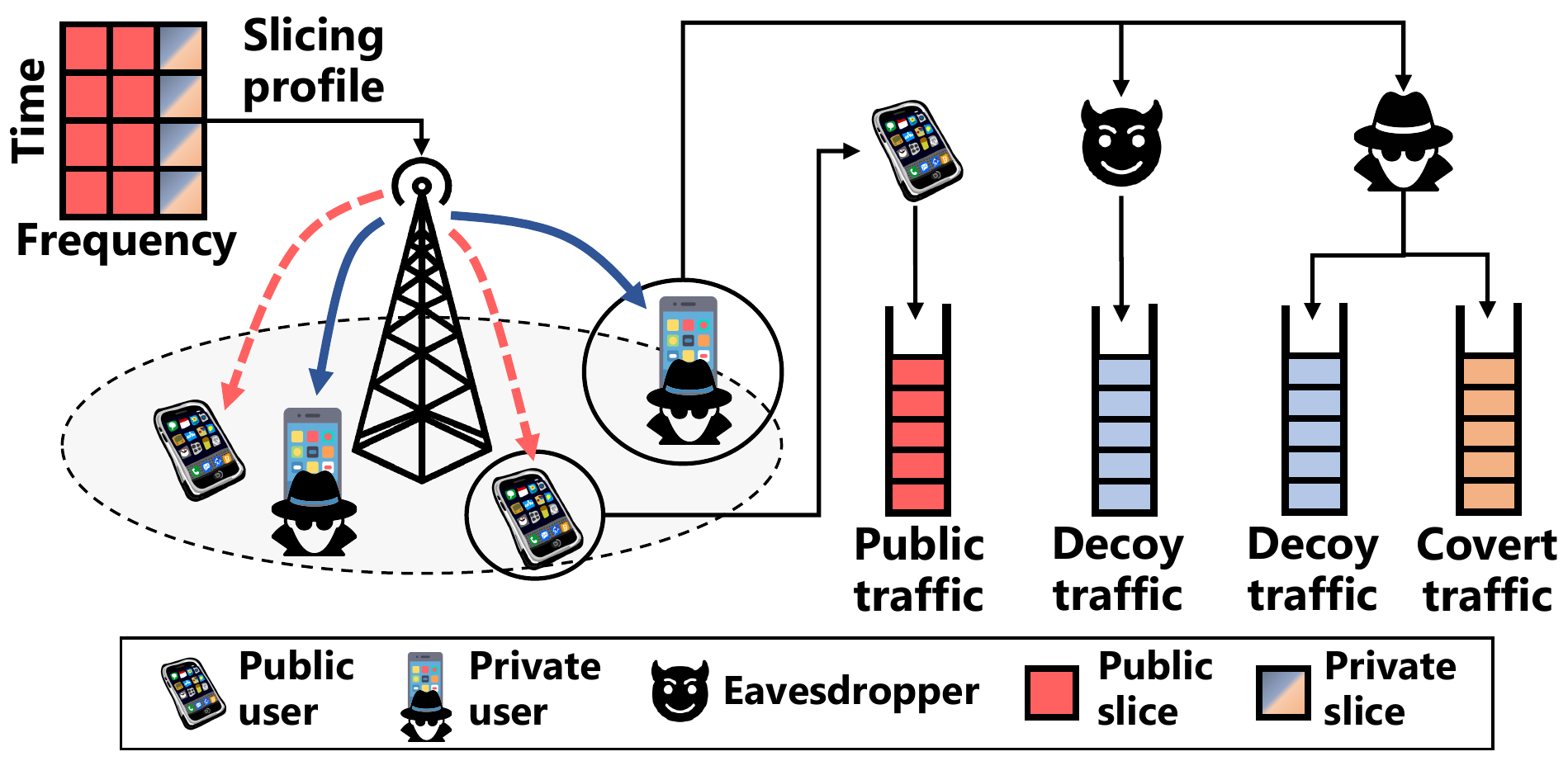}
    \end{center}
    \caption{Private Cellular Connectivity as a Service (PCCaaS).}
    \label{fig:app}
\end{figure}

The opportunities and applications of \pccaas are multifold.
%
For instance, with \pccaas law enforcement agencies could leverage the ubiquitous connectivity offered by extant cellular infrastructure and use private slices to establish undetectable communications with undercover agents in the field. 
Similarly, law enforcement could deploy tiny \ac{iot} devices as ``bugs,'' collecting audio and video content and communicating it covertly.
Such devices would pose as regular \ac{iot} sensors and conceal sensitive covert information on top of innocuous primary traffic, \eg temperature readings. 
An example of  \pccaas is shown in \fig{fig:app}.

The IP instantiates three slices whose profile is communicated to the cellular base station.
%
Two slices are public (in red).
These slices are for public users of the infrastructures (\eg cellular subscribers).
The third slice is private. 
In this slice, private users exchange data that is of non-sensitive nature (decoy traffic, in blue).
This is their primary traffic.
They also exchange \emph{covert} traffic (in orange), which is hidden into the primary data.
The challenge of \pccaas is that of fooling a malicious eavesdropper to believe that the private users are only exchanging decoy traffic, namely, in allowing the eavesdropper to capture only their primary traffic.

Clearly, no form of effective data encryption is the solution to realizing the vision of \pccaas.
First of all, not all devices have the necessary resources
to support the execution of power-hungry and computationally complex encryption algorithms.
The IoT scenario described above is a typical example.
Furthermore, encrypted traffic is still subject to jamming:
An adversarial user that is capable to detect the transmission of sensitive information could prevent its intended recipient to receive it.
The key question is therefore how to ensure that communication of sensitive data is not only secure, but also \emph{undetectable}, independently of encryption.


One of the key challenges in realizing \pccaas is that data transmitted over wireless channels cannot be easily hidden. 
To address this problem, \textit{wireless steganography} directly operates on \ac{rf} waveforms by applying ``hand-crafted'' tiny displacements to the I/Q symbols being transmitted, also known as \textit{primary} symbols~\cite{shih2017digital, dutta2012secret, grabski2013steganography, szczypiorski2010hiding, kho2007steganography, zielinska2011direct, bash2015hiding,kahn1996history}. 
While a steganographic receiver (the private user of \fig{fig:app}) can decode the covert information by translating the ``dirty'' I/Q symbols to a corresponding covert bit sequence, public users would be able to decode primary symbols only.
%

\begin{figure}[h]
    \begin{center}
        \includegraphics[width=0.95\columnwidth]{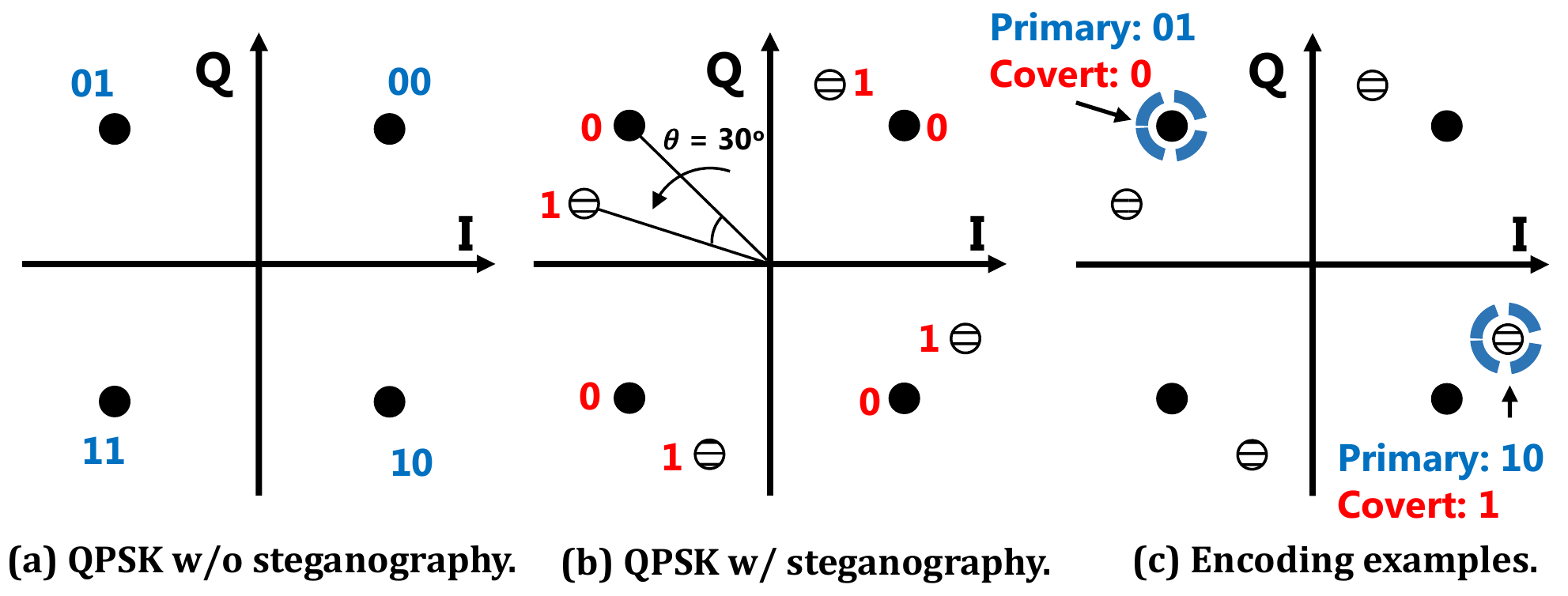}
    \caption{Wireless steganography over a QPSK modulation.}
    \label{fig:const_stego_intro}
    \end{center}
\end{figure}

\fig{fig:const_stego_intro} illustrates a practical example of wireless steganography where covert data are embedded into a QPSK-modulated signal. 

Specifically, \fig{fig:const_stego_intro}a shows the set of QPSK symbols used to form the primary message. 
Let us assume that the transmitter sends the primary symbol ``01'' to deceive adversaries, but at the same time it wants to embed a covert message in it through wireless steganography. 
To achieve this, the transmitter can rotate the phase of the I/Q symbols of \fig{fig:const_stego_intro}a by an angle~$\theta$ to send the covert bit ``1'' (\fig{fig:const_stego_intro}b), while symbols with no rotations correspond to the bit ``0'' (\fig{fig:const_stego_intro}c)~\cite{classen2015practical}.
%
(More sophisticated schemes are described in Section~\ref{sec:transmitter_design}.)


Despite recent advances,
wireless steganography has not yet found widespread application in networking. 
We believe this is because existing approaches operate only at the physical layer, which is insufficient to make \pccaas systems possible.

In this paper we leverage steganography as the core of a ready-to-use \textit{full-stack} approach to \pccaas-based networking.
Our system, that for testing purposes has been realized on open-source LTE implementations, is called \stealte to indicate the \emph{stealthy}, private nature of the networking it enables to satisfy \pccaas requirements.
\stealte achieves:

\noindent
$\bullet$ \textit{End-to-end Reliability and Security}.
We design a \textit{full-stack steganographic system} leveraging proven reliable data transfer techniques.
%
These are integrated to a \textit{steganographic mutual authentication} mechanism where legitimate parties authenticate each other before exchanging confidential information.

\noindent
$\bullet$ \textit{Adaptive Traffic Embedding}. Covert data need to be embedded over \textit{primary} traffic, which is inherently varied and unpredictable.
Clearly, a large covert data packet cannot be embedded on a small primary packet, or cannot be transmitted at all in the absence of primary traffic. 
This requires the process of embedding covert traffic to be \textit{flexible enough} to deal with such unpredictability. 
To this purpose, \stealte features a covert packet generator component 
that creates and embeds covert packets that seamlessly adapt to primary data traffic, generating ``dummy'' primary traffic on-demand, if necessary.



\noindent
$\bullet$ \textit{Standard compliance}. To successfully operate over existing cellular networks, \pccaas must adhere to standard protocol implementations of 4G/5G systems.
\stealte has been designed to seamlessly integrate with cellular systems without disrupting primary communications or affecting their performance. 

\noindent
$\bullet$ \textit{Undetectability}. A  goal of \pccaas is to make covert data communications undetectable, concealing them from eavesdroppers and jammers.
To this aim, we design a stochastic steganography scheme that embeds covert transmissions by mimicking wireless channel noise.
We show that \stealte reduces the \ac{ks} distance from the ``clean'' (i.e., without covert data) distribution by $4.8$x, improving undetectability with respect to previous solutions~\cite{doroInfocom2019Stego}.

Our LTE-compliant prototype of \stealte---the first for \pccaas-based cellular networking---has been evaluated through experiments over indoor and outdoor testbeds (including the POWDER platform from the PAWR program~\cite{breen2020powder, pawr})
on scenarios with varying parameters, including topology, traffic patterns, mobility and link ranges.
%
Our results show that, overall, the \stealte covert throughput is comparable to the primary throughput, that it minimally affects primary transmissions imposing $< 6\%$ loss of primary throughput, and that, even in challenging outdoor settings, effectively delivers covert data on links up to~$852\:\mathrm{ft}$ long.

The rest of paper is organized as follows. 
The \stealte system design is presented in Section~\ref{sec:system_design}.
Its prototype over LTE-compliant implementations is described in Section~\ref{sec:prototype}.
Section~\ref{sec:experimental_evaluation} reports results from our testbed-based experimental evaluation of \stealte.
A review of previous work on the topic is surveyed in Section~\ref{sec:related_work}.
Conclusions are drawn in Section~\ref{sec:conclusions}.

\section{S\lowercase{tea}LTE Design}
\label{sec:system_design}

In this section we describe \stealte, providing details on its \textit{covert communications} (Section~\ref{sec:covert_procedures}), \textit{transmitter} and \textit{receiver design} (sections~\ref{sec:transmitter_design} and~\ref{sec:receiver_design}), and the mechanisms to enable \textit{undetectable covert communications} (Section~\ref{sec:undetectable_covert_communications}).

\subsection{Covert Communications: Formats and Operations} 
\label{sec:covert_procedures}

%


This section describes the packet format and the operations that allow \stealte to enable \pccaas-based reliable and secure covert communications.

\subsubsection{Packet format}
\label{sec:packet_format}

\begin{figure}[h]
    \begin{center}
        \subcaptionbox{\label{fig:packet}Packet structure.}{\includegraphics[height=3.8cm]{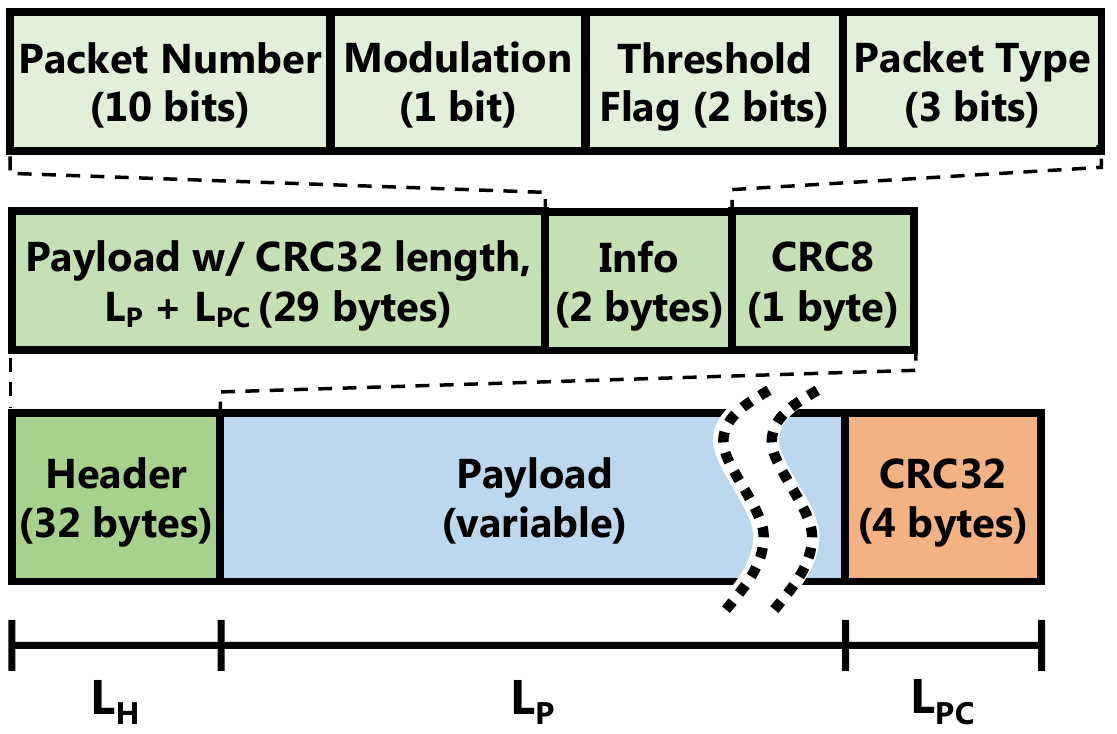}}\hspace{1pt}%
        \subcaptionbox{\label{tab:packet_types}Packet types.}{\includegraphics[height=3.8cm]{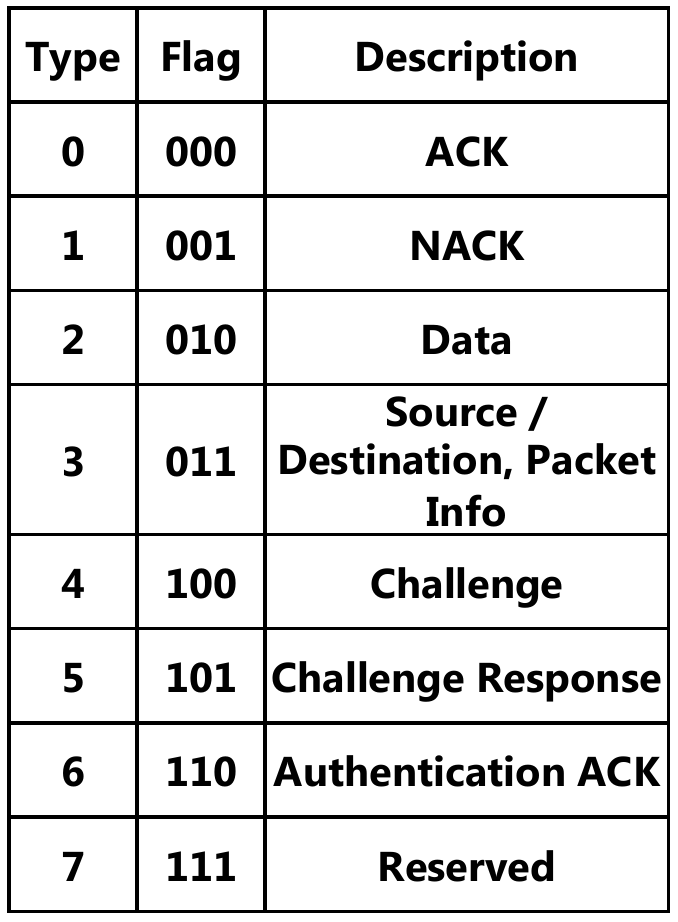}}%
    \end{center}
    \caption{\stealte covert packet structure and types.}
    \label{fig:packet_format_table}
\end{figure}

The structure of \stealte covert packets is illustrated in \fig{fig:packet}. Each packet consists of three elements: A \textit{header}, a \textit{payload} and the \emph{\ac{crc}}.


The \textbf{header} consists of~32 bytes carrying information on how to decode a received covert packet. For packet detection and demodulation,
the header is modulated through a fixed covert modulation known by the receiver.
Its structure is as follows: (i) A 29-byte field with the length of the covert payload ($L_P$) and the \ac{crc} ($L_{PC}$); (ii) a 2-byte \textit{info} field with information on how to demodulate the covert packet, and (iii) a 1-byte CRC8 field to detect errors on the header. The \textit{info} field contains:

\noindent
$\bullet$ \textit{Packet Number:} 10 bits uniquely identifying a packet, also used to request packet retransmission (Section~\ref{sec:reliable_communications}).

\noindent
$\bullet$ \textit{Modulation:} A bit flag indicating the modulation used to encode payload and \ac{crc}.
\stealte chooses between two covert modulation schemes depending on the quality of the wireless channel. This field can be extended to account for additional covert modulation schemes (see also Section~\ref{sec:adaptive_traffic_embedding}).

\noindent
$\bullet$ \textit{Threshold Flag:} 2 bits to instruct the receiver on how to demodulate and decode
covert data. This field is paramount
for our undetectability scheme (Section~\ref{sec:undetectable_covert_communications}).


\noindent
$\bullet$ \textit{Packet Type:} A 3-bit field to discern among data and control packets. The different packet types and their flags are shown in \fig{tab:packet_types}. Packet types~0 and~1 are \ack and \nack control packets sent by the receiver to give feedback on the covert transmission
(Section~\ref{sec:reliable_communications}).
Packets carrying covert data are of type~2. Packets of type~3 carry information on source and destination of covert packets and on the total number of packets in the current transmission.
Each source and destination address is encoded by~5 bytes containing the corresponding \ac{msin} (\ie the telephone number commonly used to identify mobile subscribers).
Upon receiving an uplink covert transmission, the \bs maps the destination \ac{msin} to the corresponding \ac{imsi}, which uniquely identifies the \ue.
It then relays the covert message to the receiver via a downlink covert transmission or forwards it to the \stealte \bs serving the receiver, as for regular voice traffic).
Packets of type~4,~5 and~6 are used for the mutual authentication of covert transmitters and receivers (Section~\ref{sec:mutual_authentication}).
Packet type~7 is reserved for future use.

\textbf{Payload and CRC32.}~The variable-size packet payload carries sensitive user data to be transmitted covertly. This field adapts to the size of primary packets to improve the efficiency of covert communications (Section~\ref{sec:adaptive_traffic_embedding}).
To ease reception, the length of this field is included in the packet header (\fig{fig:packet}).
The packet ends with a 4-byte CRC32 field utilized for error detection and to ensure the integrity of covert transmissions.

\subsubsection{Reliable covert communications}
\label{sec:reliable_communications}

\stealte provides built-in reliability through standard reliable data transfer mechanisms.
These include error detection, receiver-to-transmitter feedback (positive and negative acknowledgments), packet sequence numbers, timeouts, and retransmissions~\cite{KuroseR17}. 
Error detection is performed through two \acf{crc} codes: A CRC8 code is used to protect the header of the packet and a CRC32 code for the packet payload (see \fig{fig:packet}).

\subsubsection{Mutual Authentication}
\label{sec:mutual_authentication}
\stealte implements a scheme for the mutual authentication of \acp{bs} and \ues{} through covert challenge/response operations.
%
%
After standard cellular attachment procedures are completed, the \bs sends a randomly-generated \textit{challenge} to the \ue using a type~4 packet (\fig{tab:packet_types}).
Upon receiving this packet the \ue computes the \ac{hmac} from the \bs challenge and key (which has been pre-shared),
and sends the \ac{hmac} result as the \textit{challenge response} (packet of type~5). After receiving the response from the \ue, the \bs compares it with the expected \ac{hmac} result.
If the two match, the \bs considers the \ue \textit{authenticated}.
To notify the successful end of the \ue authentication procedures, the \bs sends an \textit{authentication \ack} message to the \ue (packet of type~6).
If the challenge response is not received, the \bs retransmits the challenge to the \ue.
After a certain number of unresponded attempts,
or in case of erroneous response, the \bs considers the \ue \textit{not authenticated} and will avoid any covert communication with it. 
When the \ue receives the authentication \ack from the \bs, it follows a similar procedure to \textit{authenticate} the \bs.


\subsection{Transmitter Design}
\label{sec:transmitter_design}

This section presents the main components of \stealte transmitter: The \textit{Covert Packet Generator}, the \textit{Covert Modulator}, and the \textit{Covert Embedder}.
%
In the reminder of the paper, \textit{orange-colored} blocks with dashed lines denote system components of \stealte. 
All other colors identify standard cellular components that do not require hardware or software modifications.

\begin{figure}[ht]
    \centering
    \includegraphics[width=0.95\columnwidth]{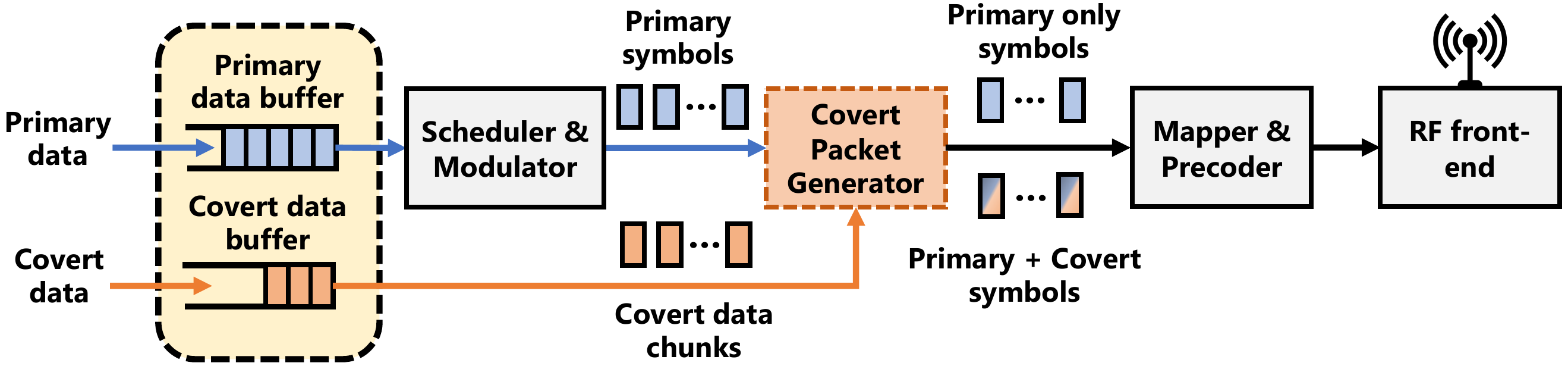}
    \caption{High-level \stealte transmitter design.}
    \label{fig:tx_design}
\end{figure}

\fig{fig:tx_design} provides a high-level overview of the building blocks of a \stealte transmitter.

Primary and covert data streams are separate and independent from one another. %
Primary data are processed through standard scheduling and signal processing procedures (\eg modulation) of the primary system. 
These data result in a sequence of primary symbols that are fed to the \stealte \textit{covert packet generator} (Section~\ref{sec:adaptive_traffic_embedding}).
After the covert symbols have been embedded in the primary symbols, they are mapped and precoded according to standard cellular procedures (Section~\ref{sec:prototype}), and transmitted through the \ac{rf} front-end.

\subsubsection{Covert Packet Generator} \label{sec:adaptive_traffic_embedding}

\begin{figure}[ht]
    \centering
    \includegraphics[width=0.95\columnwidth]{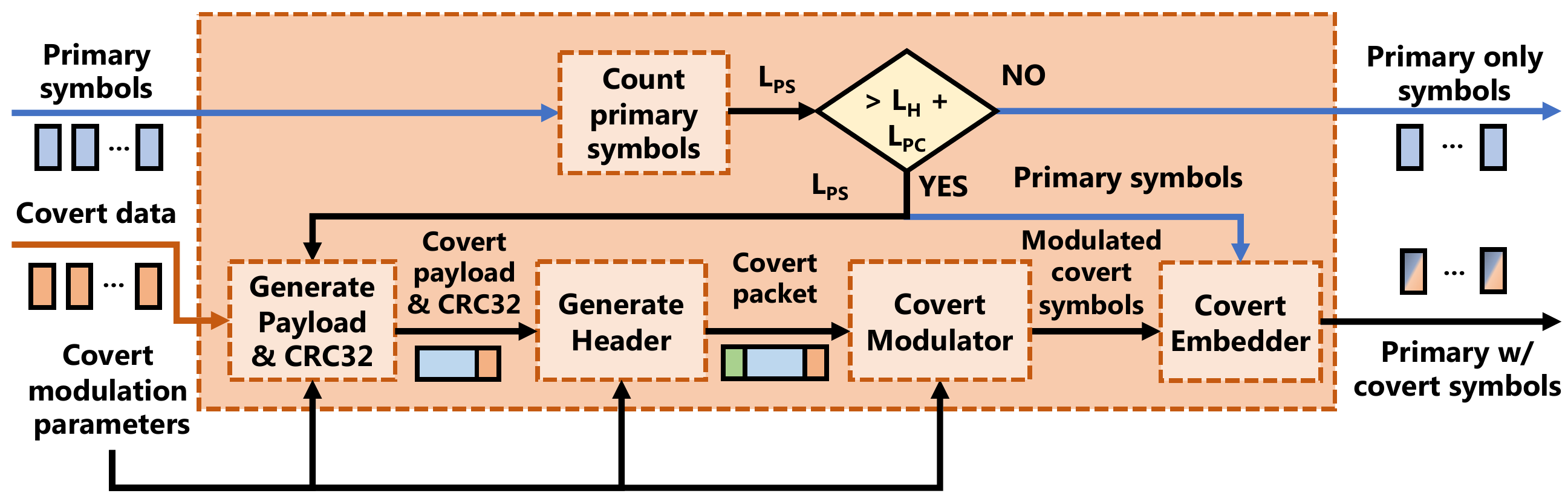}
    \caption{Covert packet generator block overview.}
    \label{fig:packet_gen}
\end{figure}

This block reads covert data from the covert data buffer, and embeds it in the modulated primary symbols. This is achieved by executing the following three steps (see \fig{fig:packet_gen}): (i) Verifying that there are enough primary symbols to embed a complete covert packet; (ii) generating covert symbols to be transmitted, and (iii) embedding them into primary ones. 

The covert packet generator starts by verifying if the number of primary symbols $L_{PS}$ is large enough to accommodate at least $L_{min}=36$ bytes, which are required for the covert packet header ($L_H=32$ bytes) and the \acs{crc}32 field ($L_{PC}=4$ bytes). 
In the positive, it generates the covert packet payload and \acs{crc}32 field. The length of the payload and of the \acs{crc}32 are included in the packet header, as described in Section~\ref{sec:packet_format}.
The packet is then modulated through the \textit{covert modulator}
according to set \textit{covert modulation parameters} (also in the header).
Finally, the resulting covert modulated symbols are embedded in the primary symbols through the \textit{covert embedder}.
If $L_{PS} \le L_{min}$ no covert data are embedded in the primary traffic. Note that the adaptive structure of the covert packets allows to embed variable size covert data on top of \textit{time-varying} and \textit{unpredictable} primary traffic. 
This feature makes \stealte transparent to primary traffic dynamics, thus enabling the integration of \stealte with any softwarized cellular system.

\begin{figure}[ht]
    \begin{center}
        \includegraphics[width=0.95\columnwidth]{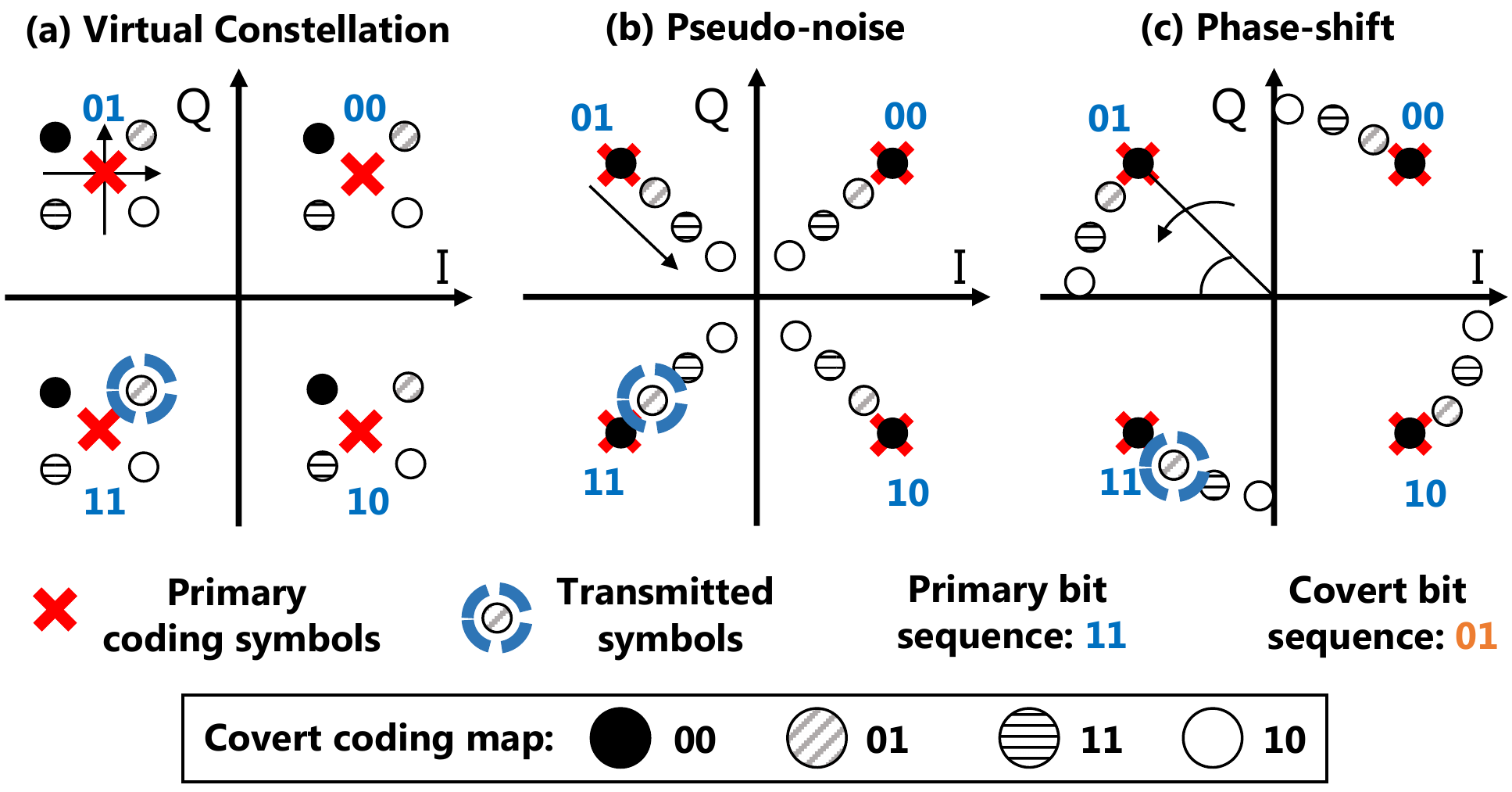}
        \vspace{-0.2cm}
    \caption{\label{fig:const_stego}Approaches to wireless steganography.\vspace{-0.1cm}}
    \end{center}
\end{figure}

\textbf{Covert Modulator.}
This block is in charge of encoding covert packets into covert symbols that can be embedded into primary transmissions (\fig{fig:packet_gen}).
Several approaches are possible for covert embedding of data through wireless steganography. 
\fig{fig:const_stego} illustrates three examples of
2 covert bits to be added on top of a primary QPSK constellation.

The first approach generates a ``dirty'' QPSK constellation around each primary symbol (\fig{fig:const_stego}a) mimicking a hierarchical constellation on top of the primary QPSK constellation~\cite{dutta2012secret}. 
The second one introduces a hierarchical \ac{ask} modulation manipulating the amplitude of the primary symbols (\fig{fig:const_stego}b) such that different amplitude values encode different covert bit sequences~\cite{doroInfocom2019Stego}. 
The third approach modifies the phase offset of the primary symbols (\fig{fig:const_stego}c) in a way that each phase rotation encodes a specific bit sequence~\cite{classen2015practical}. 
As \stealte is not tied to any specific steganographic procedure its covert modulator supports any of these approaches. 
In the following we assume that the covert modulator block implements the approach depicted in \fig{fig:const_stego}b~\cite{doroInfocom2019Stego}, which we call $M_C$-ASK, where $M_C$ is the number of symbols in the covert constellation.
The advantages of this approach include that it is robust against phase rotations introduced by fading, that it supports high-order modulation schemes (for high covert data rates), and that it can be seamlessly integrated with OFDM systems such as those used in the latest generations of cellular networks. 
The covert modulator receives the covert packets together with the set of covert modulation parameters, which specify the modulation order,
the corresponding coding map, and the packet type (\fig{fig:packet_gen}). The coding map uniquely associates covert packets (\ie bit sequences) to modulated covert symbols.
Covert symbols are, then, embedded in primary symbols through the \textit{covert embedder} block.


\textbf{Covert Embedder.}
Once covert symbols have been generated, they are embedded by the covert embedder (\fig{fig:packet_gen}).
This procedure modulates the amplitude (and phase) of the primary symbols based on the covert symbols to embed~\cite{doroInfocom2019Stego}.
The output is a sequence of primary symbols with embedded covert data. The symbols are then processed by mapping and precoding blocks and transmitted through the \ac{rf} front-end (\fig{fig:tx_design}).

\subsubsection{Downlink and Uplink Procedures}

\begin{figure}[ht]
    \begin{center}
        \subcaptionbox{\label{fig:downlink_transmitter}Downlink (DL) transmitter.}{\includegraphics[height=6.7cm]{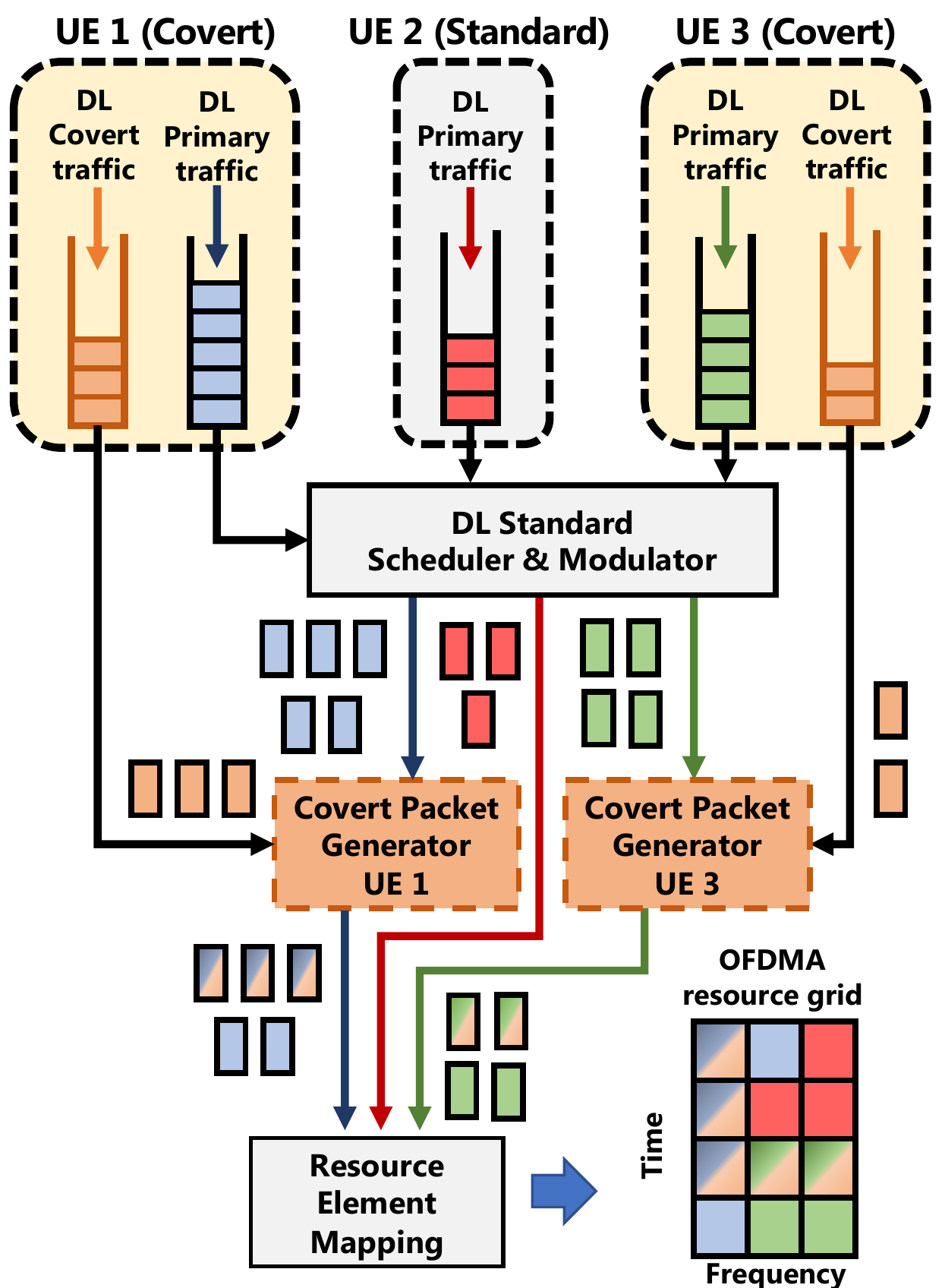}}\hspace{10pt}%
        \subcaptionbox{\label{fig:uplink_transmitter}Uplink (UL) transmitter.}{\includegraphics[height=6.7cm]{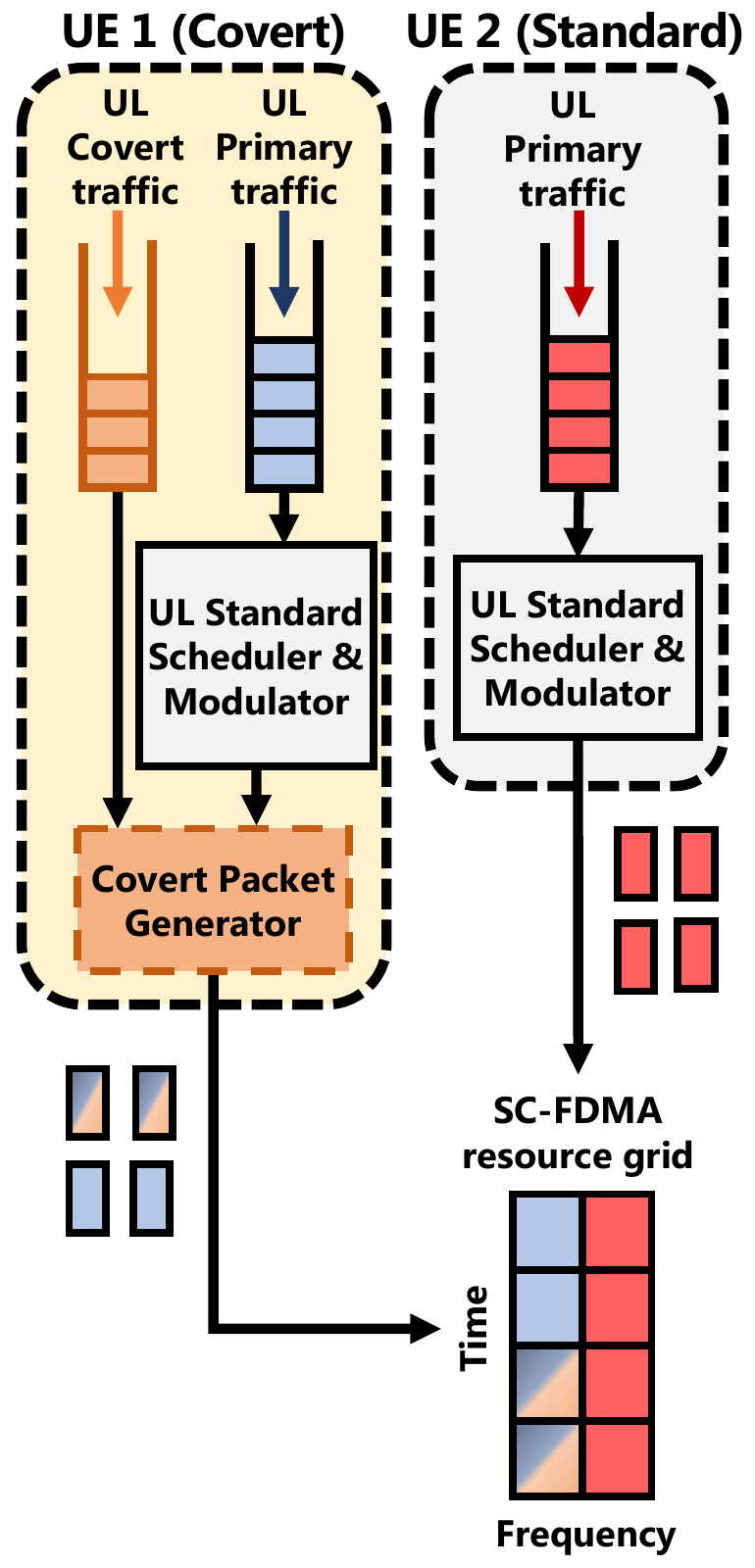}}
    \end{center}
    \vspace{-0.2cm}
    \caption{High-level \stealte downlink and uplink transmitter design.}
    \label{fig:downlink_uplink_transmitter}
\end{figure}

\stealte runs seamlessly on both downlink and uplink transmissions (\fig{fig:downlink_uplink_transmitter}), and does not depend on specific \ac{mac} strategies (\eg TDD/FDD, OFDMA/SC-FDMA). 

\fig{fig:downlink_transmitter} shows the downlink transmitter design at the \bs.
In this example, the \bs is serving three subscribers: Two covert users (\ue~1 and \ue~3),
and a standard user (\ue~2).
After scheduling the primary transmissions
through typical cellular procedures,
the \bs generates the covert packet to embed on the primary traffic of \ue~1 and \ue~3 (Section~\ref{sec:adaptive_traffic_embedding}).
Then, the data for all users is mapped on the cellular resource grid (\eg the OFDMA grid in case of LTE downlink), and transmitted.

The high-level uplink transmitter design is shown in \fig{fig:uplink_transmitter}, where two users are connected to a \stealte \bs: \ue~1 (covert user), and \ue~2 (standard user). After completing the mutual authentication procedures, \ue~1 generates and embeds the covert packets in the primary uplink traffic to send to the \bs. Then, it maps the primary uplink transmission with embedded covert data on the cellular resource grid (\eg SC-FDMA grid in case of LTE networks). On the other hand, \ue~2, which is not aware of the ongoing covert communications, schedules its uplink transmission according to standard cellular procedures.

\subsection{Receiver Design}
\label{sec:receiver_design}

\begin{figure}[ht]
    \begin{center}
        \includegraphics[width=0.95\columnwidth]{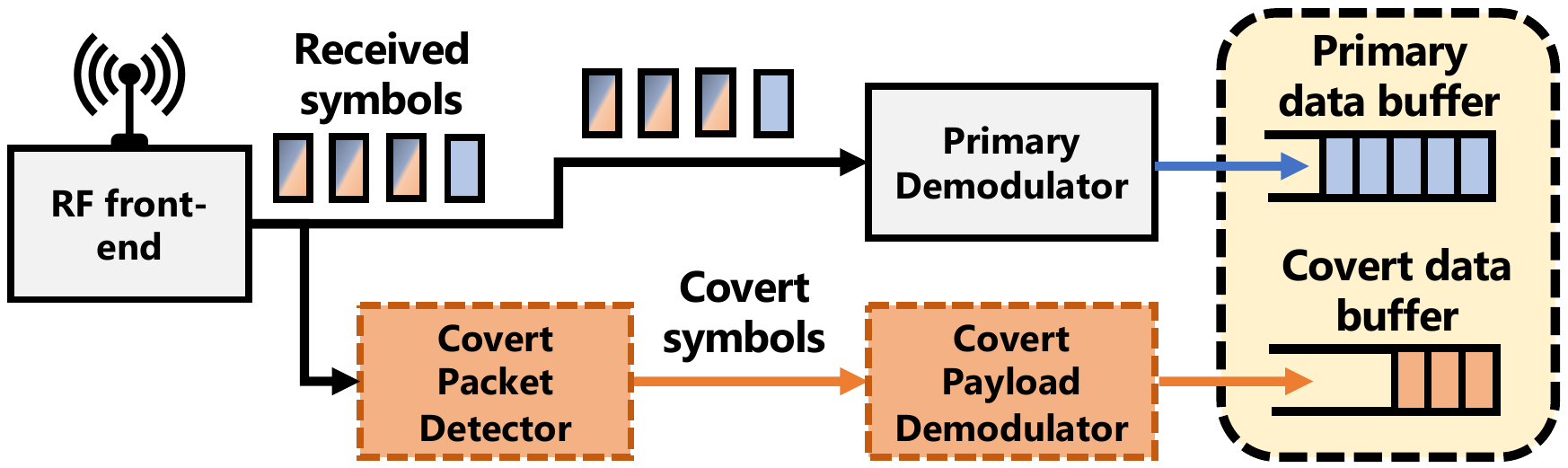}
    \end{center}
    \vspace{-0.3cm}
    \caption{High-level \stealte receiver design.}
    \label{fig:rx_design}
\end{figure}

\fig{fig:rx_design} shows the receiver design. Primary data are processed as per standard cellular procedures. 
Covert data follow a separate receive chain with two main components: The \textit{Covert Packet Detector} and the \textit{Covert Payload Demodulator}.

\subsubsection{Covert Packet Detector}
This block detects the presence of covert packets
demodulating the covert packet header
(first $L_H$ bytes of the covert packet). As mentioned in Section~\ref{sec:packet_format}, the covert header is modulated through a \bask modulation. Thus, if the \acs{crc}8 check passes (Section~\ref{sec:reliable_communications}), the receiver assumes a covert packet has been received.

Upon detecting a covert packet, the covert packet detector reads the length $L_P + L_{PC}$ of the covert payload and \acs{crc}32 fields,
the packet number and the modulation parameters 
in the info field of the header (\fig{fig:packet}).
Finally, it extracts the symbols corresponding to the encoded $L_P + L_{PC}$ bytes of the covert packet, that will be demodulated by the \textit{covert demodulator} block of Section~\ref{sec:covert_demodulator}. 

\subsubsection{Covert Demodulator}
\label{sec:covert_demodulator}
This block extracts the encoded covert information from each packet. As shown in \fig{fig:packet}, covert modulation parameters necessary to demodulate covert packets, such as employed covert modulation, packet length and the packet type are specified in the header. This way, the demodulator block can reconstruct the decoding map and use it to demodulate the received symbols into covert data (\eg bit sequence).
Since all the covert modulation parameters are specified in the packet header, no further interaction is required between transmitter and receiver.
Section~\ref{sec:undetectable_covert_communications} (below) shows that this approach also enables time-varying coding/decoding mappings that make covert transmissions undetectable and secure against eavesdroppers. 
Finally, the received \acs{crc}32 value is checked (Section~\ref{sec:reliable_communications}). If the check passes, the data are saved, otherwise a retransmission will be requested.

\subsection{Undetectable Covert Communications}
\label{sec:undetectable_covert_communications}

\begin{figure}[ht]
    \begin{center}
        \subcaptionbox{\label{fig:iq_pdf_cdf_a}\acs{pdf} w/o covert.}{\includegraphics[width=0.48\columnwidth]{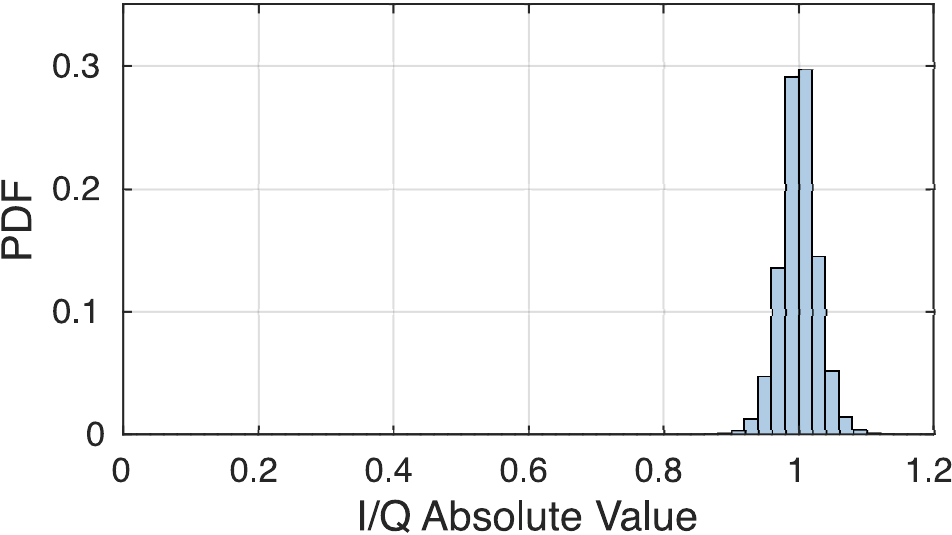}}\hspace{1pt}\hspace{0.01\columnwidth}%
        \subcaptionbox{\label{fig:iq_pdf_cdf_b}\acs{pdf} w/ fixed covert.}{\includegraphics[width=0.48\columnwidth]{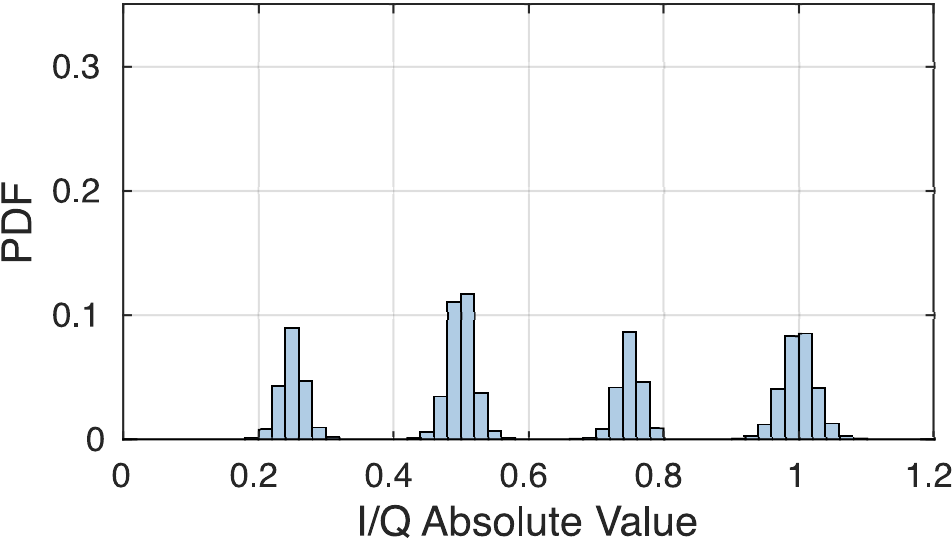}}\hspace{1pt}\\\vspace{10pt}%
        \subcaptionbox{\label{fig:iq_pdf_cdf_c}\acs{pdf} w/ undetectable covert.}{\includegraphics[width=0.48\columnwidth]{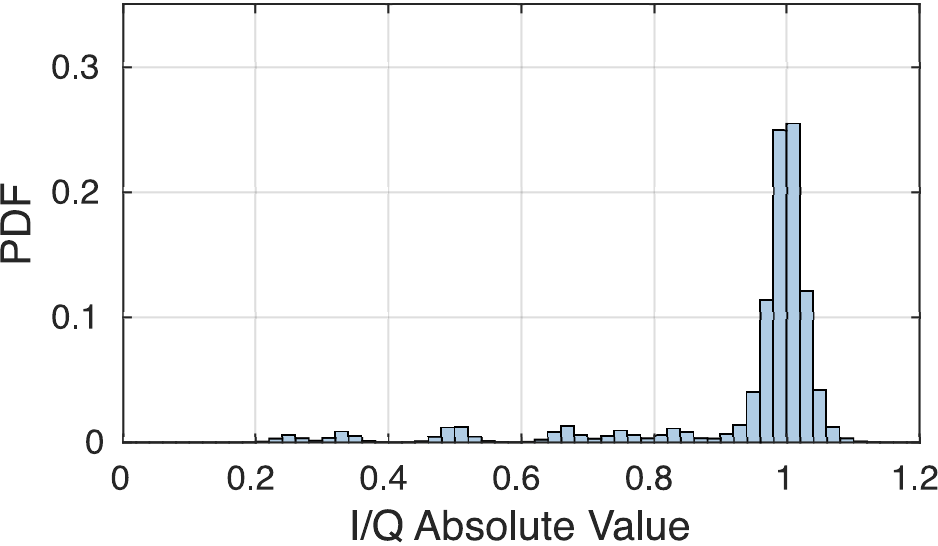}}\hspace{0.01\columnwidth}%
        \subcaptionbox{\label{fig:iq_pdf_cdf_d}\acs{cdf} and K-S distance.}{\includegraphics[width=0.48\columnwidth]{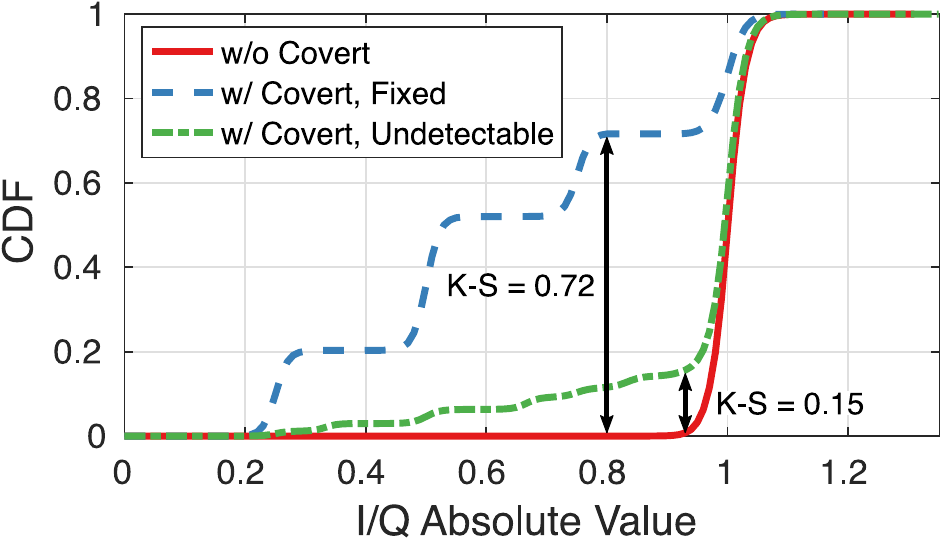}}
    \end{center}
    \caption{\acs{pdf} and \acs{cdf} of received I/Q samples.}
    \label{fig:iq_pdf_cdf}
\end{figure}

Steganography is not immune from attacks.
%
For instance, through steganalysis~\cite{xia2014steganalysis} an eavesdropper may analyze the statistical properties of captured I/Q samples and infer 
the presence of a covert slice.
For example, let us consider the case of primary QPSK transmissions where \stealte embeds covert data through a \qask covert modulation~\cite{doroInfocom2019Stego}. \fig{fig:iq_pdf_cdf} shows the \ac{pdf} of the I/Q samples captured through the testbed described in Section~\ref{sec:exp_setup} in three different cases: Primary-only transmissions are shown in \fig{fig:iq_pdf_cdf_a}; primary with \textit{fixed}, i.e., detectable, \qask covert transmissions in \fig{fig:iq_pdf_cdf_b}, and primary with \stealte \textit{undetectable} covert transmissions in \fig{fig:iq_pdf_cdf_c}.
We also show the \ac{cdf} of all cases in \fig{fig:iq_pdf_cdf_d}.
The \acf{ks} distance is also shown to measure the similarity of the \acp{cdf}: The smaller the distance, the better.

\fig{fig:iq_pdf_cdf_a} shows the \ac{pdf} of the absolute value of the captured primary I/Q samples without any covert transmission. As expected, the \ac{pdf} assumes a Gaussian distribution with mean $1$ due to noise and fading.
%
%
Fig.~\ref{fig:iq_pdf_cdf_b} shows the \ac{pdf} when covert data are embedded through a \qask covert modulation~\cite{doroInfocom2019Stego}.
Comparing \fig{fig:iq_pdf_cdf_b} with the primary-only case of \fig{fig:iq_pdf_cdf_a}, we notice that the absolute value of the captured samples no longer exhibits a Gaussian \ac{pdf} centered around $1$, but multiple bell-shaped Gaussian curves centered at~$0.25$, $0.5$, $0.75$, and~$1$. This is also illustrated in \fig{fig:iq_pdf_cdf_d} where the \ac{cdf} of the I/Q samples resembles a step function with a K-S distance with the primary-only case equal to~$0.72$.
Such statistical behavior is not surprising: This result is inherited from the \qask covert scheme in \fig{fig:const_stego}b, whose operation results in 4 possible covert points per primary symbol with amplitude equal to $0.25$, $0.5$, $0.75$, and~$1$. 
Steganalysis can easily identify such an abnormal statistical pattern, thus revealing the ongoing covert transmissions to the eavesdropper.  
For steganographic communications to be undetectable, they must statistically behave like primary ones, which is possible by reducing their K-S distance. 

\begin{figure}[h]
    \begin{center}
        \subcaptionbox{\label{fig:undetect_const_a}Flag = 01.}{\includegraphics[height=2.7cm]{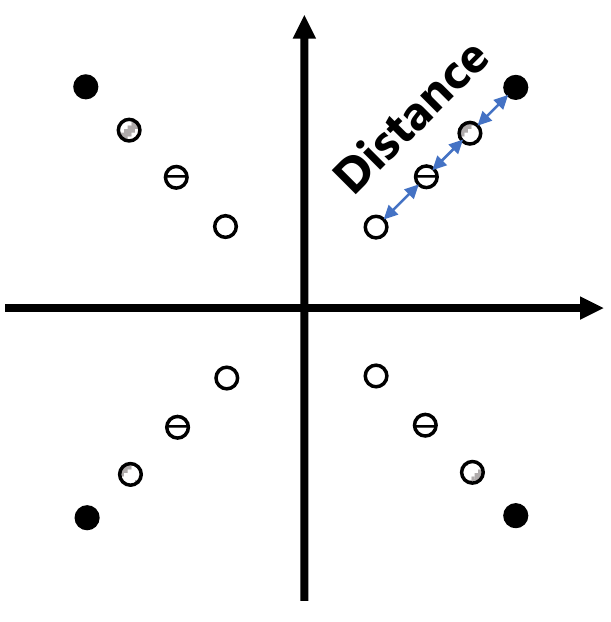}}\hspace{10pt}%
        \subcaptionbox{\label{fig:undetect_const_b}Flag = 11.}{\includegraphics[height=2.7cm]{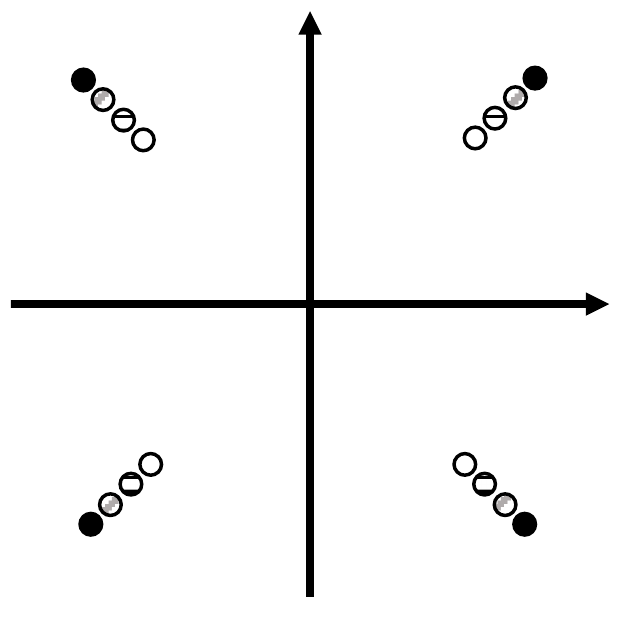}}\hspace{10pt}%
        \subcaptionbox{\label{fig:undetect_const_c}Thresholds.}{\includegraphics[height=2.7cm]{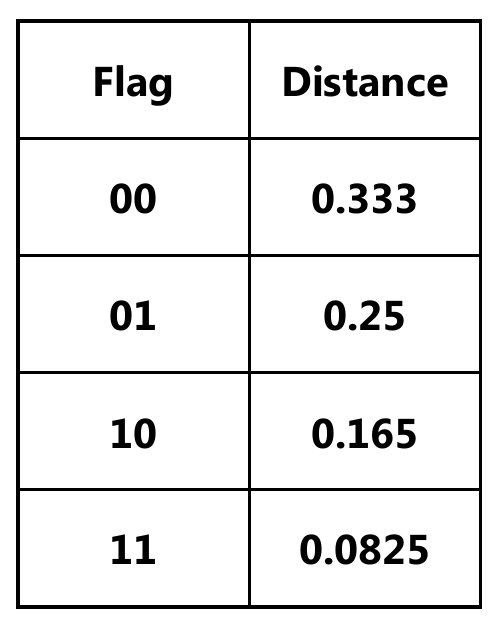}}
    \end{center}
    \caption{Examples of covert \qask constellations with different distances and threshold flag values.}
    \label{fig:undetect_const}
\end{figure}

For this reason, \stealte implements a mechanism that mimics I/Q displacements introduced by channel noise by randomizing the covert embedding procedures. 
Rather than utilizing a \textit{fixed distance} between covert symbols (as done in~\cite{doroInfocom2019Stego}), \stealte \textit{randomly changes} this distance,
providing the \textit{first-of-its-kind undetectability mechanism}.
This process is illustrated in \fig{fig:undetect_const},
where we show 4 possible configurations of a \qask covert constellation.

To decode covert ASK messages, the receiver must be aware of the distance between covert symbols. As a consequence, randomizing the transmitted covert constellation
could potentially undermine the receiver's covert demodulation procedures. 
To overcome this problem, \stealte packets carry a \textit{threshold flag} field (see Section~\ref{sec:packet_format}) instructing the receiver on the covert constellation used by the transmitter. The value of this flag is changed on a per-packet basis, thus reducing the probability of successful steganalysis attacks. In our current implementation,
this field consists of 2 bits encoding the 4 different distance configurations shown in \fig{fig:undetect_const_c}.
However, it is straightforward to increase the size of this field to further enhance undetectability.

The effectiveness of our approach is depicted in \fig{fig:iq_pdf_cdf_c}, where we show the \ac{pdf} of \stealte undetectable covert messages. The absolute value of the captured I/Q samples is centered around $1$, while the remaining small peaks are hardly distinguishable from the noise of the wireless channel. The same behavior can also be observed in the \ac{cdf} of the primary I/Q samples shown in \fig{fig:iq_pdf_cdf_d}.
When compared with the \ac{cdf} of the primary-only LTE transmissions (solid line), the \ac{cdf} of the covert signal not only does not show the steps observed with fixed displacements (dashed line), but it also results in a $4.8$x shorter K-S distance. 

\section{S\lowercase{tea}LTE Prototype}
\label{sec:prototype}

We prototyped \stealte on \ac{cots} NI USRPs B210 and X310 \acp{sdr}.
Our implementation is based on the LTE-compliant srsLTE open-source software, which offers protocol stack implementations for LTE base stations (\enbs), \ues{}, and core network~\cite{gomez2016srslte}. 
We remark that as \stealte follows a software-defined approach, it is not bound to LTE technology, and it can be easily extended to future 5G-and-beyond cellular networks.

\begin{figure}[ht]
    \centering
    \includegraphics[width=0.95\columnwidth]{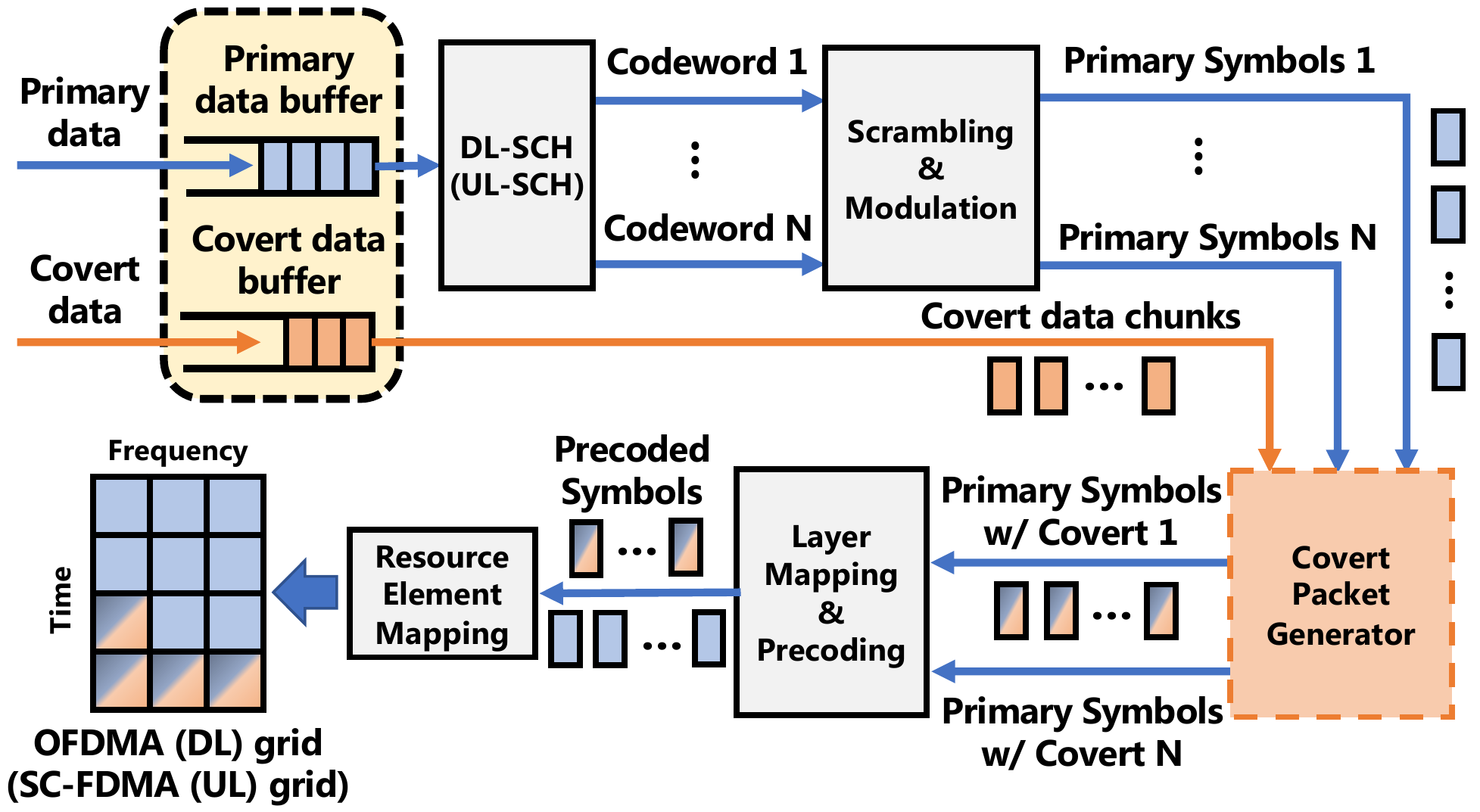}
    \caption{The \stealte covert transmitter.}
    \label{fig:pdsch_pusch_prototype}
\end{figure}

We extended srsLTE to allow \stealte to \textit{embed}, \textit{encode}, and \textit{decode} covert data on the downlink and uplink LTE primary traffic.
Specifically, we enhanced the \ac{pdsch} and \ac{pusch} procedures at the PHY-layer.
The \ac{pdsch} carries the downlink data sent by the \enb to the \ues, and the random access response messages if the \ac{pdsch} is mapped to the \ac{rach}.
The \ac{pusch} carries the uplink data that the \ues transmit to the \enb, and ACKs and NACKs for \ac{pdsch} data.
%
\fig{fig:pdsch_pusch_prototype} depicts the modified structure of the \stealte covert transmitter, \ie the \ac{pdsch} at the \enb downlink side or the \ac{pusch} at the \ue uplink side.

When there is primary data to transmit (either in downlink to the \ue or uplink to the \enb), this is converted into \textit{codewords} through the \ac{dlsch} (or the \ac{ulsch}), which performs transport channel encoding operations.
Then, the resulting codewords are \textit{scrambled} and \textit{modulated} into \textit{primary symbols} (for illustration purposes, in our experiments we adopt a QPSK modulation for the primary traffic that carries covert data).
After these operations the \stealte \textit{covert packet generator} (Section~\ref{sec:adaptive_traffic_embedding}) modifies the amplitude of the resulting primary symbols, thus applying a second (covert) modulation to them. 
This way, the \textit{covert data chunks} are embedded on the primary symbols.
%
At this point, the complex-modulated \textit{primary symbols with embedded covert} are mapped into layers for spatial multiplexing, and then precoded, as per LTE specifications. 
Finally, the resulting \textit{precoded symbols} are mapped into resource elements to be transmitted via OFDMA (downlink), or SC-FDMA (uplink).


\section{Experimental Evaluation}
\label{sec:experimental_evaluation}

We report results from experimental campaigns for evaluating the performance of \stealte.
%
Setups are described in Section~\ref{sec:exp_setup}.
Results are shown and discussed in Section~\ref{sec:results}.

\subsection{Experimental Setup}
\label{sec:exp_setup}

\noindent
We evaluated \stealte on indoor and outdoor testbeds.

\begin{figure}[ht]
    \begin{center}
        \subcaptionbox{\label{fig:experiment_setup_static}Static scenario.}{\hspace{1pt}\includegraphics[height=4cm]{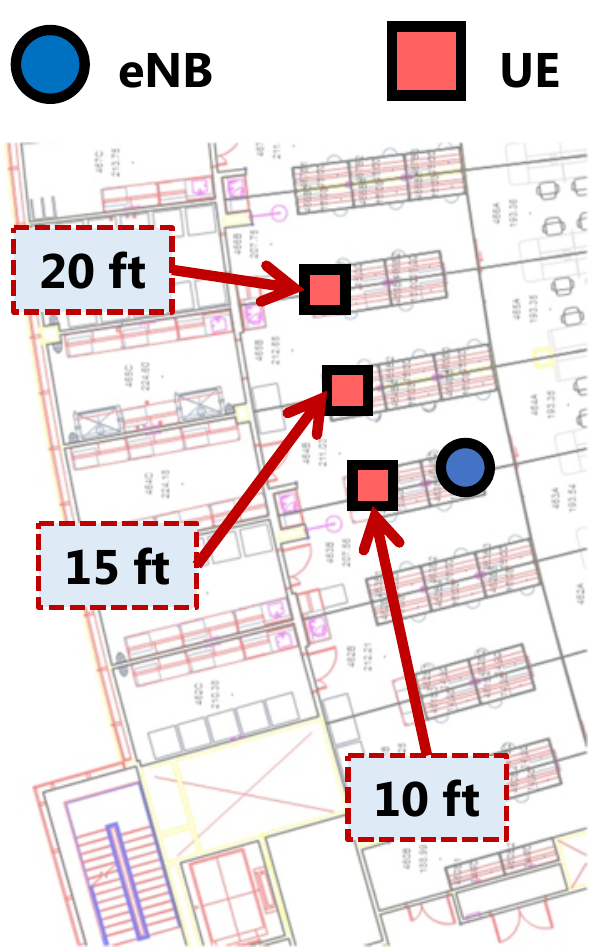}}\hspace{10pt}%
        \subcaptionbox{\label{fig:experiment_setup_dynamic}Dynamic scenario.}{\includegraphics[height=4cm]{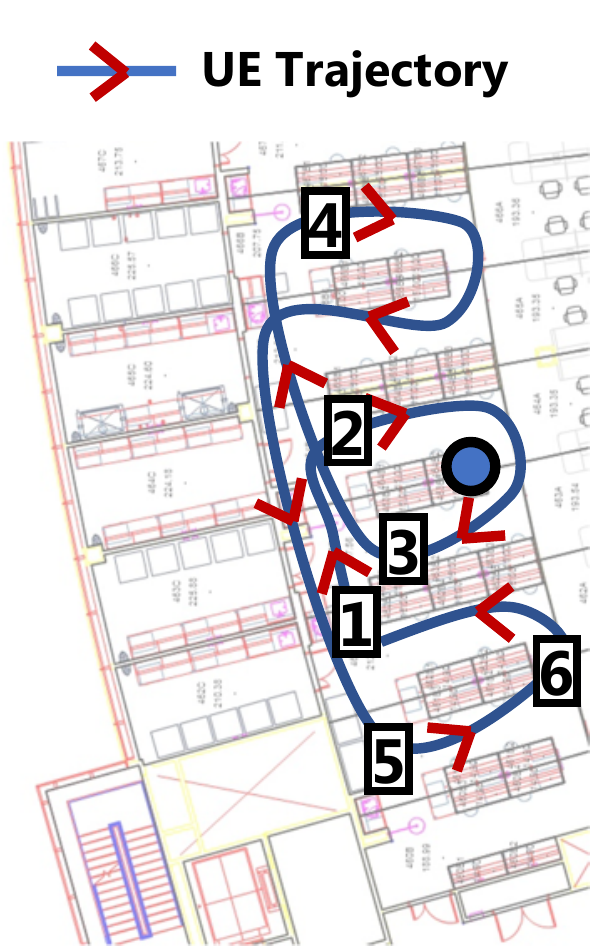}}\hspace{8pt}%
        \subcaptionbox{\label{fig:experiment_setup_cots}\pccaas scenario.}{\includegraphics[height=4cm]{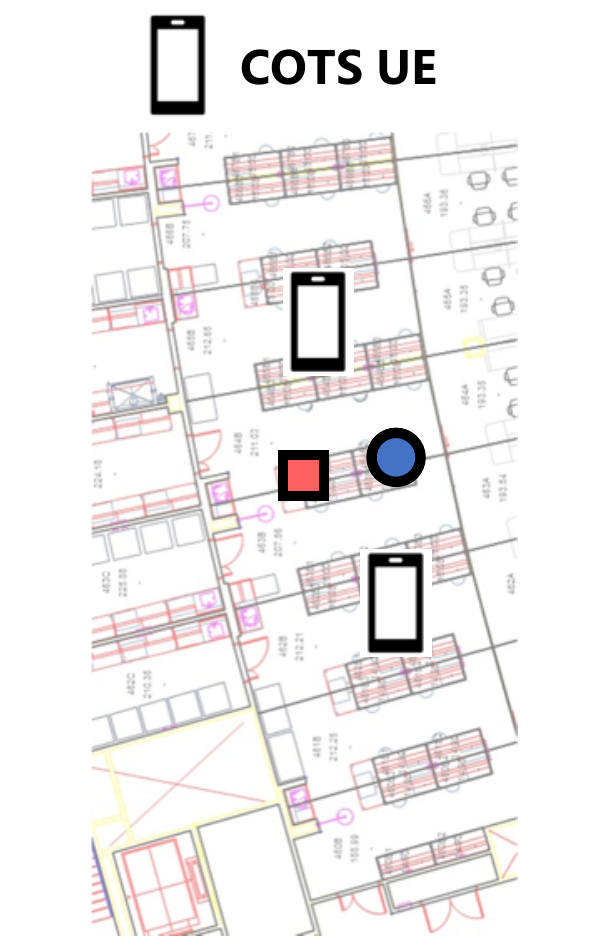}}
    \end{center}
    \caption{Indoor testbed setup and experiment configuration on the Arena testbed~\cite{bertizzolo2020arena}.}
    \label{fig:testbed}
\end{figure}

\emph{Indoor scenarios}.
For our indoor experiments we leveraged Arena, an indoor ceiling testbed covering an area of $2240\:\mathrm{ft^2}$~\cite{bertizzolo2020arena}.
We instantiate LTE-compliant \enbs and \ues on NI USRP X310 \acp{sdr}, USRP B210 and \ac{cots} Xiaomi Redmi Go smartphones.
In all configurations the \enb uses a $10\:\mathrm{MHz}$ channel bandwidth corresponding to $50$~\acp{prb}.
These devices are used in the indoor testbed configurations depicted in \fig{fig:testbed}. 

\noindent
The \emph{static scenario} comprises one \enb and three \ues that are statically placed $10\:\mathrm{ft}$, $15\:\mathrm{ft}$ and $20\:\mathrm{ft}$ away from the \enb (\fig{fig:experiment_setup_static}).
In this configuration all devices are USRPs X310.
In this scenario \enb and \ue exchange primary traffic generated through \textit{iperf3}~\cite{iperf3}, a software tool for network performance evaluation with TCP and UDP traffic.
For experiments with UDP traffic the bitrate varies in the set $\{0.75, 2.5, 5\}\:\mathrm{Mbps}$.
We also use traffic generated by a user-initiated speed test.

\noindent
The \emph{dynamic scenario} is made up of the \enb and one \ue traveling a distance of $190\:\mathrm{ft}$ around the \enb as illustrated in \fig{fig:experiment_setup_dynamic}.
Measurements are collected at six different location in the \ue journey.
In this configuration the \enb is a USRP X310 and the mobile device is a USRP B210.
Primary traffic is generated through the \textit{ping} software utility, which uses \ac{icmp} echo request and reply messages.

\noindent
The \emph{\pccaas scenario} comprises two slices, one public and one private, with one \enb (USRP X310), one \ue using the private slice for covert traffic (USRP X310), and two \ac{cots} Xiaomi Redmi Go smartphones transmitting data over the public slice.
All users are statically positioned as in \fig{fig:experiment_setup_cots} within $10\:\mathrm{ft}$ from the \enb.
In this scenario, primary traffic concerns YouTube videos streamed by the users.

\emph{Outdoor scenario}.
For outdoor, long-range testing we ported \stealte to the \ac{pawr} \ac{powder} platform~\cite{pawr, breen2020powder}. 
We use the NR version of srsLTE to instantiate one outdoor~5G base station (\gnb) and one \ue. 
The \gnb is located on the rooftop of a~$95\:\mathrm{ft}$-tall building and is realized by a USRP X310. 
The \ue is statically positioned at ground-level. 
It is implemented through a USRP B210.
The distance between \gnb and \ue is $852\:\mathrm{ft}$.
%
%
The \gnb uses a $3\:\mathrm{MHz}$-bandwidth ($15$~\acp{prb}).
Covert data are embedded through a \bask modulation.
Primary traffic between \gnb and \ue is generated through the \textit{ping} utility.

In all scenarios covert data are images and text files.

%
%
%
%
%
%
%

\subsection{Experimental results}
\label{sec:results}

In this section we report the results of the performance evaluation of \stealte in each of the considered scenarios.
For each scenario, we describe the investigated metrics and their relevance, and illustrate the corresponding experimental results.
Plots include 95\% confidence intervals (not shown if $< 1\%$).

\subsubsection{Indoor static scenario}
\label{sec:staticSce}

\begin{figure}[ht]
    \begin{center}
        \subcaptionbox{\label{fig:dl_tcp_covert_throughput}Throughput.}{\includegraphics[height=3.5cm]{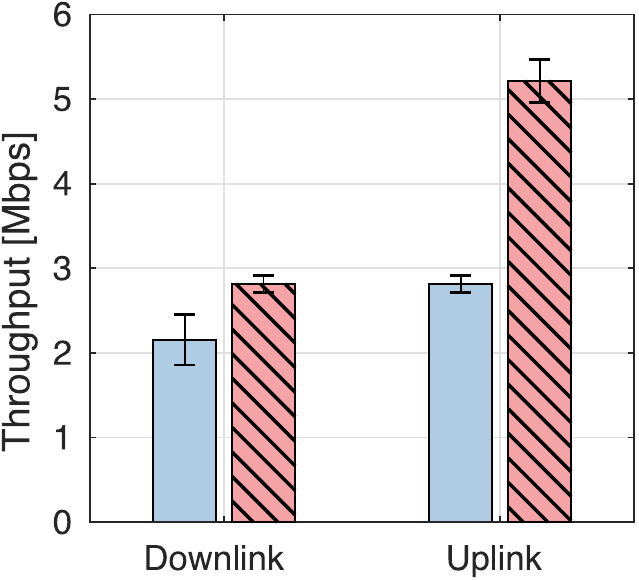}}\hspace{0.05\columnwidth}%
        \subcaptionbox{\label{fig:dl_tcp_covert_retx}Retransmissions.}{\includegraphics[height=3.5cm]{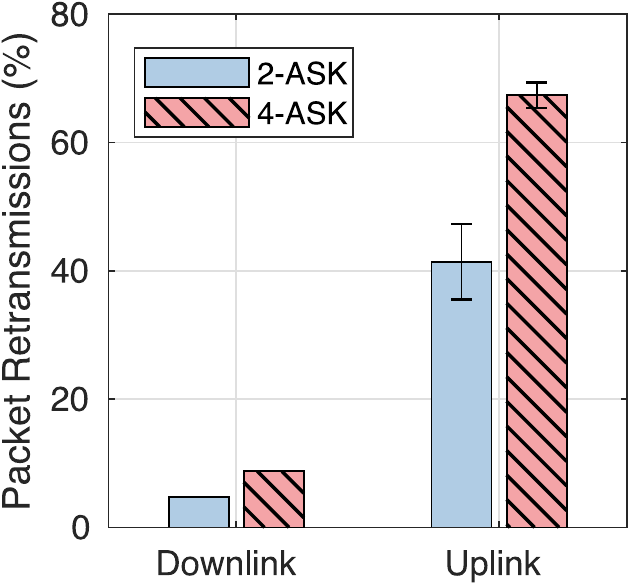}}
    \end{center}
    \caption{Downlink and uplink covert performance with TCP primary traffic for different covert modulations.}
    \label{fig:dl_tcp_covert}
\end{figure}

In the static scenario of \fig{fig:experiment_setup_static} we start by measuring the performance of \stealte to deliver covert data by investigating throughout (data delivery over time) and the percentage of packets that needs to be retransmitted.
\fig{fig:dl_tcp_covert} shows the downlink and uplink covert throughput and retransmissions for both \bask and \qask covert modulations under TCP primary traffic.

Being of higher order, \qask obtains a higher throughput than \bask (\fig{fig:dl_tcp_covert_throughput}). 
However, this comes at the cost of transmission errors, leading to a higher percentage of covert packet retransmissions (\fig{fig:dl_tcp_covert_retx}).
This is because of the higher resilience to errors of the \bask modulation, which requires~38\% less retransmissions than the \qask case (uplink).

\begin{figure}[ht]
    \begin{center}
        \subcaptionbox{\label{fig:dl_udp_covert_throughput}Throughput.}{\includegraphics[height=3.5cm]{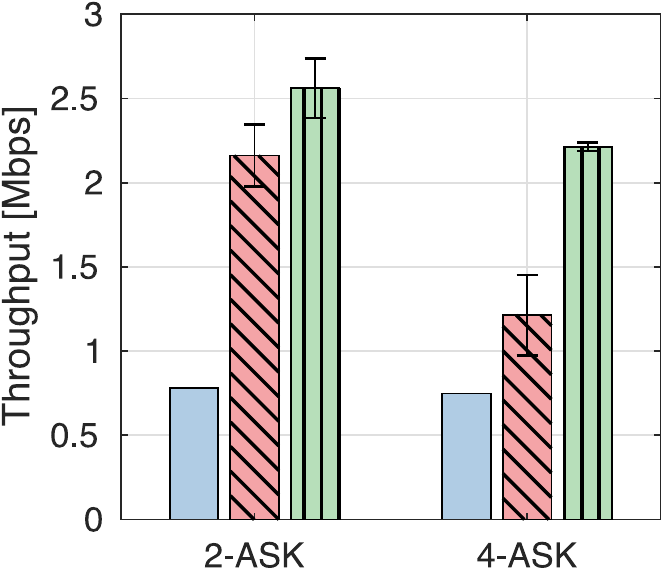}}\hspace{0.05\columnwidth}%
        \subcaptionbox{\label{fig:dl_udp_covert_retx}Retransmissions.}{\includegraphics[height=3.5cm]{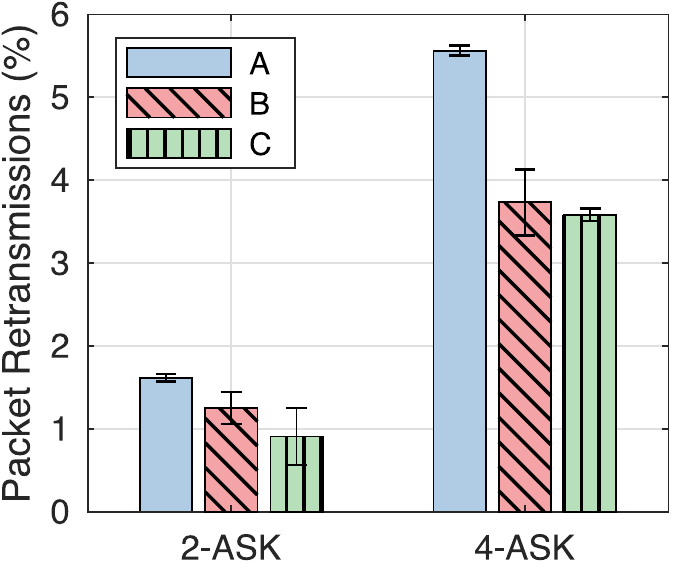}}
    \end{center}
    \vspace{-0.1cm}
    \caption{Downlink covert performance with UDP primary traffic for different traffic profiles and covert modulations.}
    \label{fig:dl_udp_covert}
\end{figure}

\fig{fig:dl_udp_covert} depicts the covert throughput (\fig{fig:dl_udp_covert_throughput}) and percentage of covert packet retransmissions (\fig{fig:dl_udp_covert_retx}) under UDP primary traffic. The \textit{iperf3} UDP bitrate was set as follows: (A)~$0.75\:\mathrm{Mbps}$; (B)~$2.5\:\mathrm{Mbps}$, and (C)~$5\:\mathrm{Mbps}$.

Differently from TCP traffic (see downlink performance in \fig{fig:dl_tcp_covert_throughput}), the covert throughput is higher when data are embedded through a \bask modulation. 
This is because, unlike TCP, UDP streams data at the specified bitrate, as it does not implement reliable data transfer.
This behavior can be further observed in \fig{fig:dl_udp_covert_retx}, which shows a much higher (close to $3$x) number of covert retransmissions in case of \qask modulation.

\begin{figure}[h]
    \begin{center}
        \subcaptionbox{\label{fig:distance_covert_throughput}Throughput.}{\includegraphics[height=3.5cm]{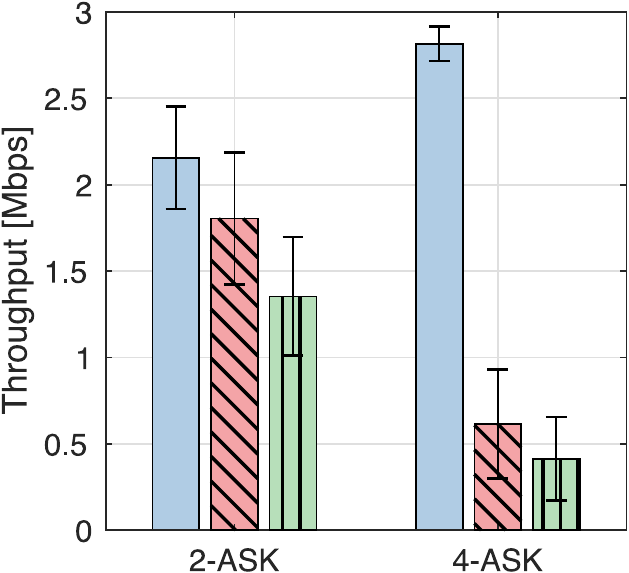}}\hspace{0.05\columnwidth}%
        \subcaptionbox{\label{fig:distance_covert_retx}Retransmissions.}{\includegraphics[height=3.5cm]{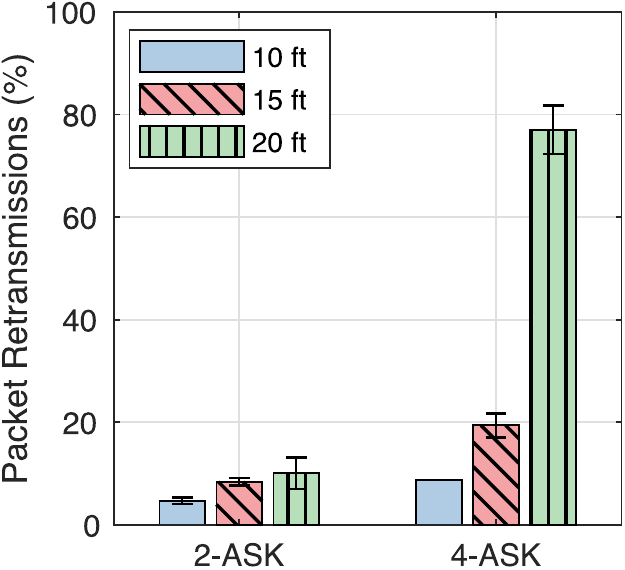}}
    \end{center}
    \vspace{-0.1cm}
    \caption{Downlink covert performance with TCP primary traffic for different covert modulations and distances between \enb and \ue.}
    \label{fig:distance_covert}
\end{figure}

Figures~\ref{fig:distance_covert_throughput} and~\ref{fig:distance_covert_retx} show the covert throughput and the percentage of packet retransmissions when TCP primary traffic is exchanged between \enb and \ue, as a function of their distance (\fig{fig:experiment_setup_static}).
We notice that for short distances (\ie $10\:\mathrm{ft}$) \qask provides the highest performance.
However, as the distance increases ($\geq 15\:\mathrm{ft}$), modulating the covert message through \bask yields a better performance, both for throughput and retransmissions. 
This is because of the higher robustness to errors of this modulation compared to the \qask modulation.

\begin{figure}[ht]
    \begin{center}
        \subcaptionbox{\label{fig:speedtest_primary_throughput}Downlink throughput.}{\includegraphics[width=0.48\columnwidth]{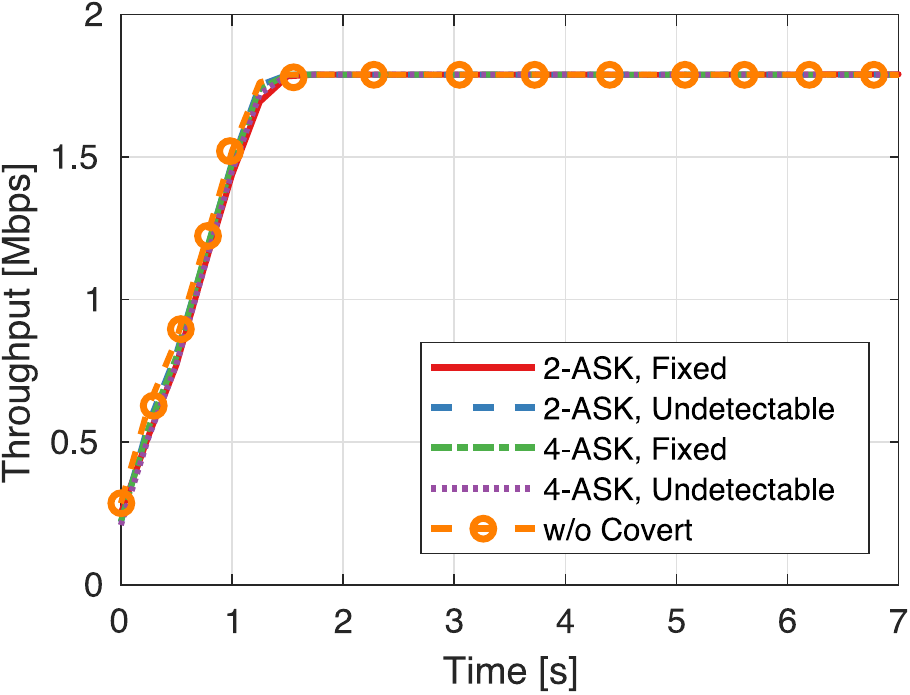}}\hspace{0.01\columnwidth}%
        \subcaptionbox{\label{fig:speedtest_primary_buffer}Downlink buffer size.}{\includegraphics[width=0.48\columnwidth]{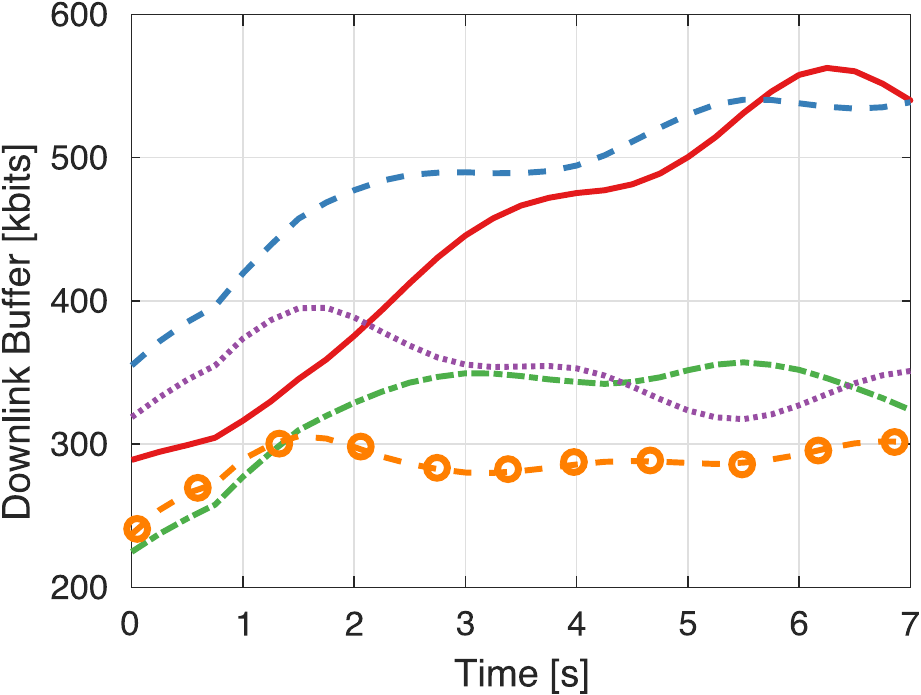}}\\%
        \vspace{10pt}\subcaptionbox{\label{fig:speedtest_primary_swarm}Downlink buffer and \acs{sinr}.}{\includegraphics[width=0.8\columnwidth]{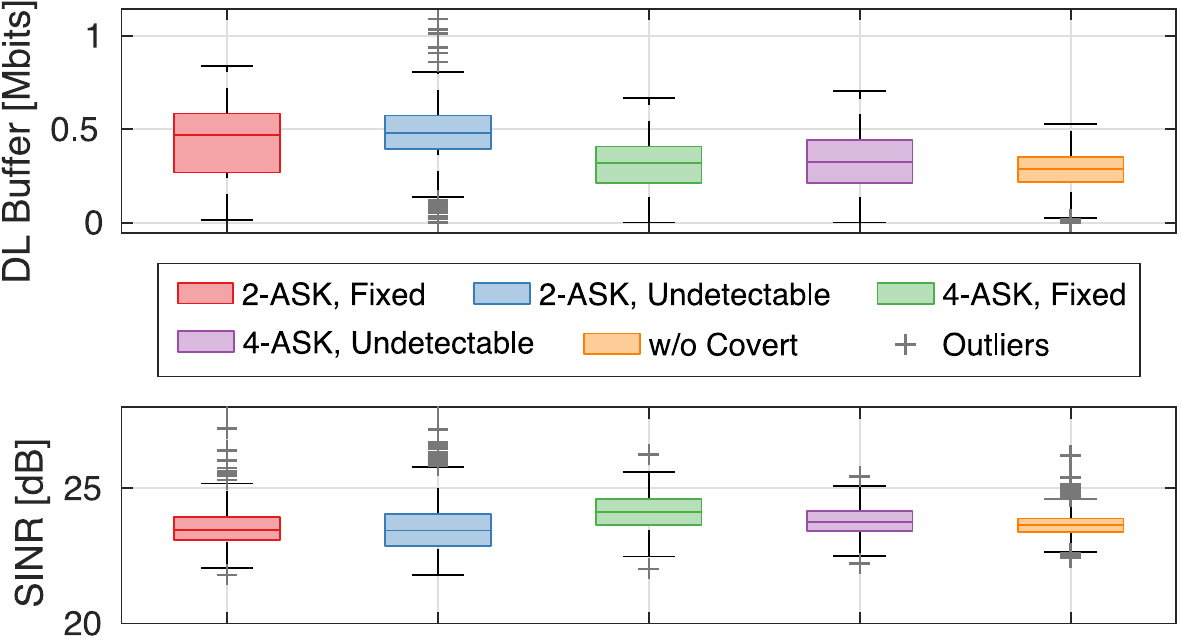}}
    \end{center}
    \vspace{-0.2cm}
    \caption{Impact of \stealte on speed test primary traffic.}
    \label{fig:speedtest}
\end{figure}

We now investigate the impact of the undetectability schemes presented in Section~\ref{sec:undetectable_covert_communications} on the performance of primary transmissions.
Their effectiveness in concealing covert data has been shown in \fig{fig:iq_pdf_cdf}.
Here we show results for metrics that indicate that these schemes do not have a significant impact over the quality of transmission and on channel quality: Throughput, the number of bytes that are to be transmitted in the downlink and the \ac{sinr}.
The results shown in \fig{fig:speedtest} refer to \ue-generated traffic according to a speed test application at~$1.8\:\mathrm{Mbps}$.

Figures~\ref{fig:speedtest_primary_throughput} and~\ref{fig:speedtest_primary_buffer} show the downlink primary throughput and transmission buffer size averaged over~10 independent experiments.
\fig{fig:speedtest_primary_throughput} clearly indicates that embedding covert messages---both fixed and undetectable---does not noticeably affect primary traffic. 
In \fig{fig:speedtest_primary_buffer} we notice a slight increase of the size of the downlink buffer queue when covert communications happen, especially for the \bask undetectability schemes.
This suggests that a higher number of retransmissions is needed for the \enb to deliver the primary traffic to the \ues, with an impact on the number of resources that the \enb uses to communicate to the users. 
Yet again, in this scenario, this does not translate in a noticeable degradation of the throughput.

\fig{fig:speedtest_primary_swarm} shows the distribution of the downlink buffer size (top) and the \ac{sinr} measured by the user (bottom).
These two results show that the statistical distributions of these two metrics do not vary significantly in the case when \stealte is used (independently of the undetectability scheme and modulation used) and the case when it is not.
Particularly, the average difference among the downlink buffer size in the two cases never exceeds $106.5\:\mathrm{kbits}$.
Also, the difference among SINR is within $0.15\:\mathrm{dB}$, indicating that embedding covert data does not increase noise perceptively. 
%

\subsubsection{Indoor dynamic scenario}
\label{sec:dynamicSce}

\begin{figure}[ht]
    \begin{center}
        \subcaptionbox{\label{fig:mobility_covert_throughput}Throughput.}{\includegraphics[height=3.5cm]{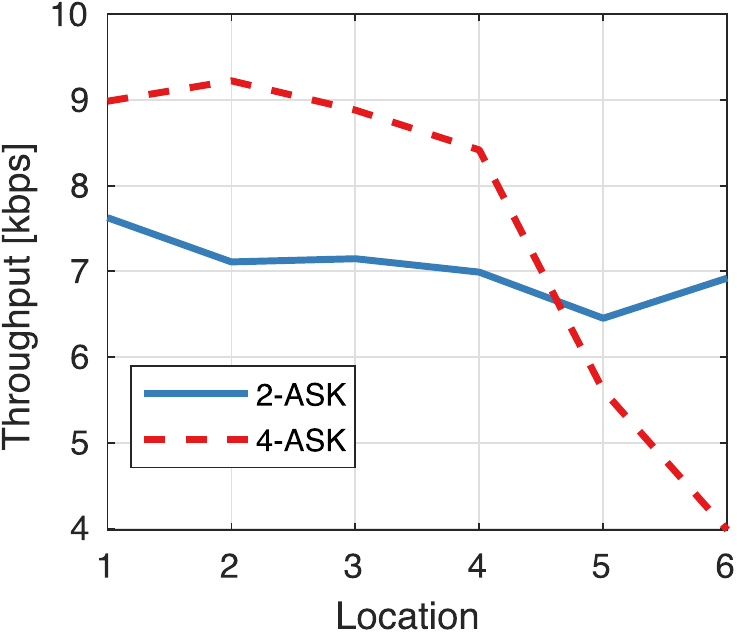}}\hspace{0.05\columnwidth}%
        \subcaptionbox{\label{fig:mobility_covert_retx}Retransmissions.}{\includegraphics[height=3.5cm]{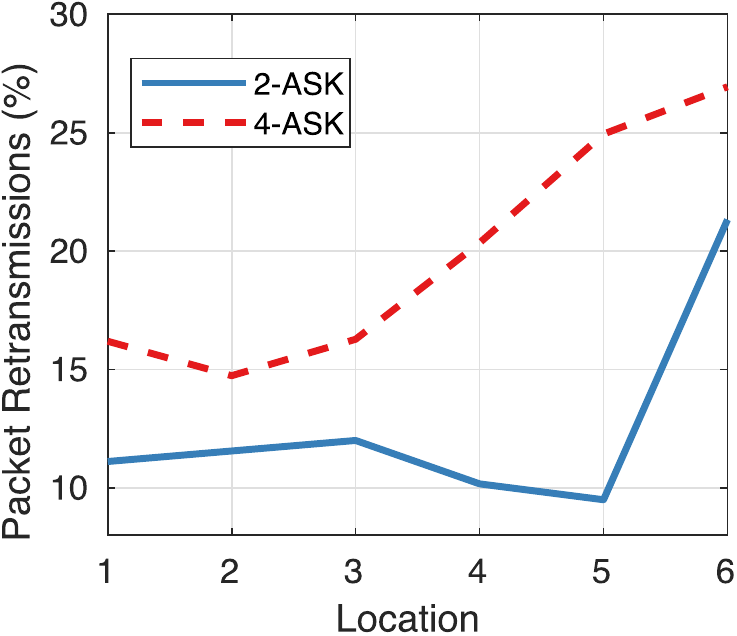}}
    \end{center}
    \vspace{-0.1cm}
    \caption{Covert performance with \acs{icmp} echo reply primary traffic for different covert modulations in presence of \ue mobility.}
    \label{fig:mobility_covert}
\end{figure}

%
The throughput and retransmission performance of \stealte for the different user locations of \fig{fig:experiment_setup_dynamic} and covert modulations is shown in \fig{fig:mobility_covert}.

As the distance between \enb and \ue increases (as for locations~4,~5 and~6---\fig{fig:experiment_setup_dynamic}), the covert throughput (\fig{fig:mobility_covert_throughput}) for both \bask and \qask modulations decreases, while the percentage of packet retransmissions increases (\fig{fig:mobility_covert_retx}).
As noticed before, in general, \bask outperforms \qask because of its higher robustness to channel impairments, which are more noticeable at larger distances.
Despite the performance degradation, \stealte still manages to enable secret communications between covert transmitter and receiver in presence of mobility.

\subsubsection{Indoor \pccaas scenario}
\label{sec:slicing}

\begin{figure}[ht]
    \begin{center}
        \subcaptionbox{\label{fig:slice_a_throughput}Allocation~A, throughput.}{\includegraphics[width=0.45\columnwidth]{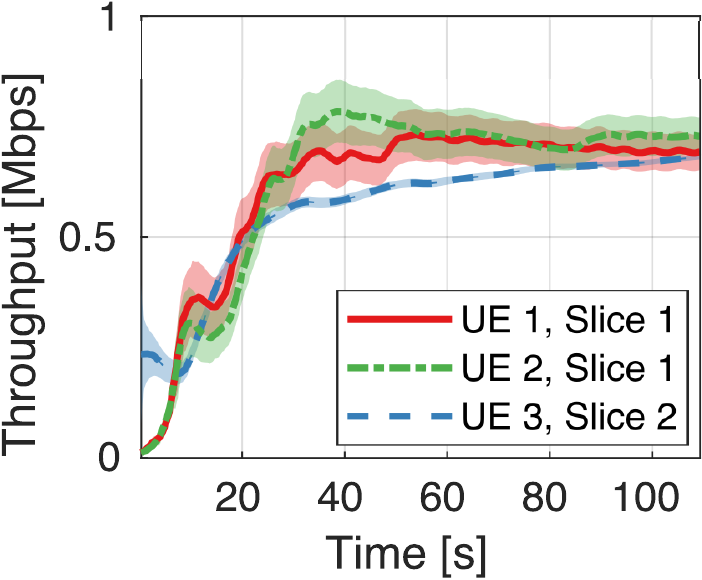}}\hspace{0.04\columnwidth}%
        \subcaptionbox{\label{fig:slice_a_errors}Allocation~A, packet errors.}{\includegraphics[width=0.45\columnwidth]{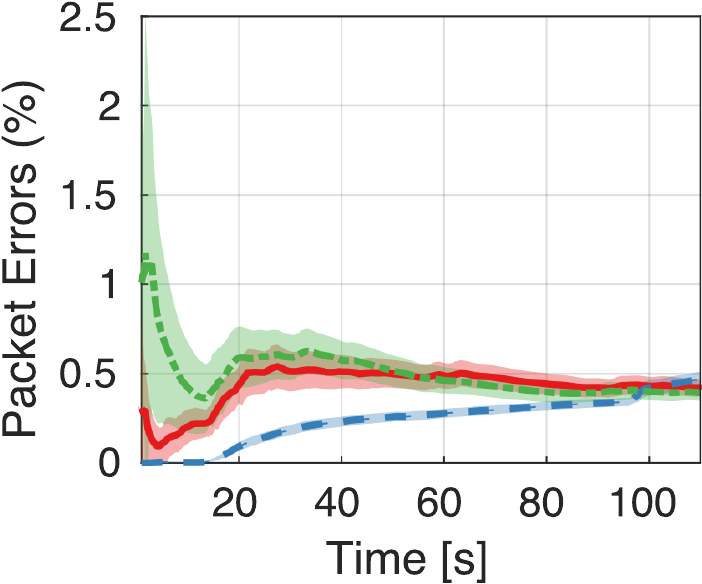}}\\\vspace{10pt}%
        \subcaptionbox{\label{fig:slice_b_throughput}Allocation~B, throughput.}{\includegraphics[width=0.45\columnwidth]{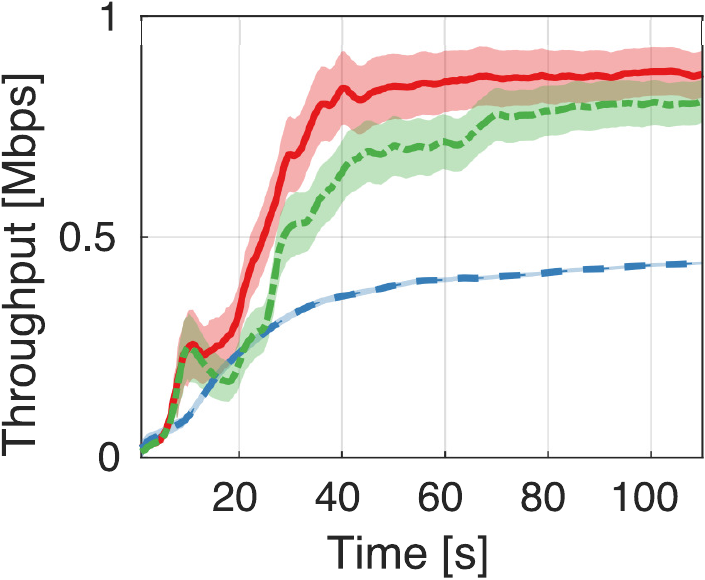}}\hspace{0.04\columnwidth}%
        \subcaptionbox{\label{fig:slice_b_errors}Allocation~B, packet errors.}{\includegraphics[width=0.45\columnwidth]{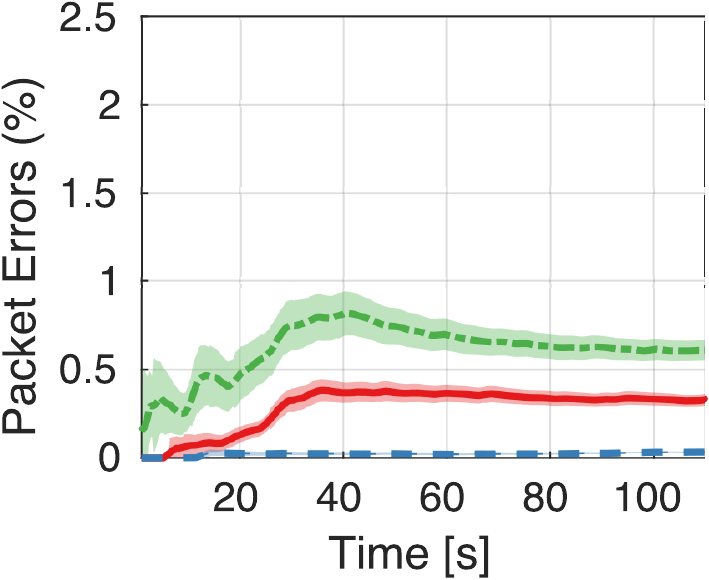}}
    \end{center}
    \caption{Throughput and percentage of packet errors on primary traffic for different resource allocations of the private and standard slices.}
    \label{fig:slicing}
\end{figure}

We now set to investigate the effectiveness of \stealte for instantiating private network slices on a shared cellular infrastructure, which serves the need of those critical applications requiring rapid instantiation of private and secure cellular networks.
In this set of experiments, we instantiate two slices: A primary-only slice (Slice~1) serving covert-agnostic \ues~1 and~2 (for which we use smartphones), and a \stealte private slice (Slice~2) for covert communications between \ue~3 and the \enb (both USRPs X310).
The covert data of \ue~3 is embedded through \qask modulation.
We consider two different network slicing allocations: (A)~Spectrum resources are evenly split among the two slices, and (B)~70\% of the resources are allocated to Slice~1 and 30\% to Slice~2.
For these scenarios we investigate the primary throughput and the percentage of packets erroneously received for both slicing configurations when all users stream videos from YouTube. 
Results are shown in \fig{fig:slicing}.

We notice that in both allocations, the primary throughput of all users is not impacted by the presence of covert communications.
The percentage of packet errors (Figures~\ref{fig:slice_a_errors} and~\ref{fig:slice_b_errors}) is negligible in both allocations (\ie below $0.5\%$ on average, with a peak of $2.5\%$).
As a result, Figures~\ref{fig:slice_a_throughput} and~\ref{fig:slice_b_throughput} show that the throughput level achieved by all users is enough for rate-demanding applications, thus confirming the low impact of \stealte on primary communications.

\subsubsection{Outdoor scenario}
\label{sec:powder}

\begin{figure}[ht]
    \begin{center}
        \subcaptionbox{\label{fig:powder_covert}Covert.}{\includegraphics[width=0.48\columnwidth]{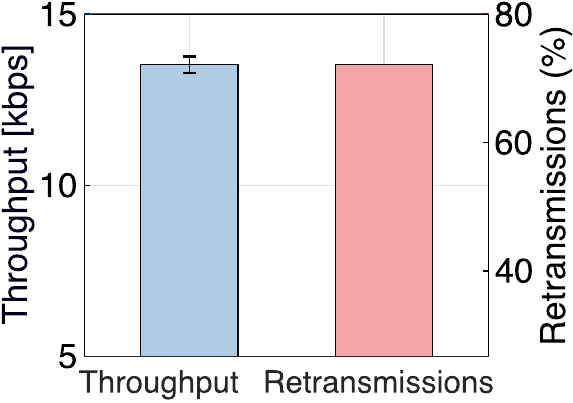}}\hspace{0.03\columnwidth}%
        \subcaptionbox{\label{fig:powder_primary}Primary.}{\includegraphics[width=0.48\columnwidth]{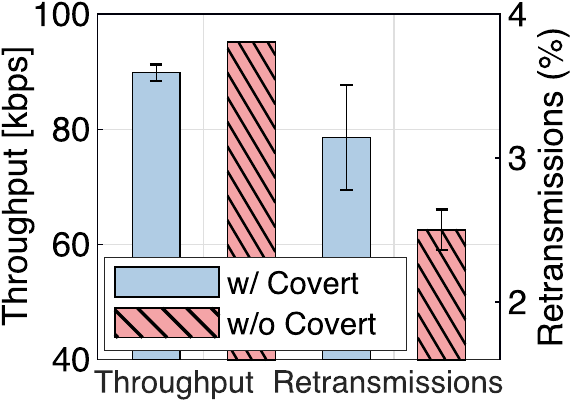}}%
    \end{center}
    \vspace{-0.2cm}
    \caption{Long-range experiments 
    on the POWDER platform.}
    \label{fig:powder}
\end{figure}

\stealte has been tested also on a long-rage link ($> 850$~ft) in the outdoor scenario provided by the \ac{powder} platform~\cite{breen2020powder}. 
Results on throughput and retransmission percentage are shown in \fig{fig:powder}.
Specifically, \fig{fig:powder_covert} shows downlink covert throughput and retransmissions and \fig{fig:powder_primary} depicts the same metrics for the primary traffic (with and without covert). 

\fig{fig:powder_covert} shows that \stealte is capable of delivering covert data despite the severe path-loss, multi-path and fading conditions experienced over the long-range link, thus demonstrating its suitability for traditional (mostly outdoors) cellular applications.
As for the indoor results above, \fig{fig:powder_primary} confirms that embedding covert data into primary traffic has negligible effect on its throughput performance.
The primary throughput degradation over the long-range link is merely $5.59\%$, whereas the packet retransmission percentage increases from $2.6\%$ to $3.2\%$ ($0.6\%$ increase) when covert traffic is embedded.

\section{Related Work}\label{sec:related_work}

Wireless steganography has been frequently used for covert communications among parties.
Differently from approaches where covert data are embedded in the packet control fields (\eg checksum~\cite{liu2014data}, flags~\cite{murdoch2005embedding, ahsan2002practical}, and padding fields~\cite{grabska2014steganography}, among others~\cite{martins2010steganography, mehta2008steganography}), wireless steganography introduces tiny displacements in the I/Q constellation plane that can be controlled to encode covert information. 
Typical methods include frequency/phase shifts~\cite{grabski2013steganography,classen2015practical}, I/Q imbalance~\cite{sankhe2019impairment}, superimposing noisy constellations~\cite{doroInfocom2019Stego,cao2018wireless,kumar2014phy,dutta2012secret} or training and preamble sequences manipulations~\cite{classen2015practical,hijaz2010exploiting}. %
These approaches, however, lack reliability as they are prone to demodulation errors and rarely support long-range communications, quintessential for many communication systems. 

These reliability issues have been partially addressed at the higher layers of the protocol stack. 
Hamdaqa and Tahvildari describe a steganographic system for Voice-over-IP (VoIP) that encodes covert information by carefully delaying packet transmissions~\cite{hamdaqa2011relack}.
Although this approach is highly reliable and undetectable, it operates over large temporal windows, which considerably limits the achievable covert rate.
Nain and Rajalakshmi develop a steganographic communication system that hides information over chip sequences of IEEE 802.15.4 networks integrating error-coding techniques to mitigate errors~\cite{nain2016reliable}. Although being reliable, this method only achieves low transmission rates.
%
Overall, previous solutions either achieve \textit{low covert throughput}, or are highly detectable through steganalysis~\cite{grabski2013network,xia2014steganalysis},
or lack practical implementations demonstrating their effectiveness and feasibility.

\stealte is the missing answer to the high throughput, reliability, and undetectability requirements of \pccaas applications characterized by mobility, time-varying channels and large distances.
As a reliable end-to-end steganographic system \stealte: (i) Achieves high throughput through wireless steganography; (ii)  provides reliable and channel-resilient communications through a combination of error-coding, retransmissions and adaptive covert modulation schemes, and (iii) can be seamlessly integrated in 3GPP-compliant cellular systems. 
%

\section{Conclusions}
\label{sec:conclusions}

This paper proposes \stealte, the first practical \pccaas-enabling system for softwarized cellular networks.
%
\stealte supports reliable, undetectable, high-throughput, and long-range covert communications.
We have prototyped \stealte and implemented it on LTE-compliant testbeds, including indoor settings and the \acs{pawr} \acs{powder} platform (ourdoors).
We have extensively evaluated the performance of \stealte under diverse traffic profiles, distance and mobility patterns, highlighting the feasibility of undetectable transmissions and their negligible impact on primary traffic.
Results over the multiplicity of the considered scenarios show that, in the vast majority of our experiments \stealte achieves a throughput of covert traffic that is at least 90\% of the throughput of the primary traffic, affecting the latter negligibly ($< 6\%$ loss).

\balance

\footnotesize
\bibliographystyle{IEEEtran}
\bibliography{biblio}

\end{document}